\documentstyle[12pt]{article}

\newcommand{\rew}[4]{{\it {#1}\,}{\bf {#2}\,}{#3\,}{#4}}
\newcommand{\nuc}{Nucl.Phys.}
\newcommand{\lett}{Phys.Lett.}

\newcommand{\ho}{\begin{equation}}
\newcommand{\la}{\end{equation}}

\begin{document}

\renewcommand{\theequation}{\thesection.\arabic{equation}}
\newcommand{\Section}[1]{\section{#1}\setcounter{equation}{0}}

\begin{titlepage}

\begin{raggedleft}
  arch-ive/9504012\\
  CSIC-IMAFF-15-1995\\
  April 1995\\
\end{raggedleft}

\begin{center}
\huge
{ \bf Fokker-Planck approach to }\\
\vspace{6pt}
{ \bf quantum lattice Hamiltonians}\\
\vspace{6pt}

\vspace{2 true  cm}

\large
{Fernando Jim{\'e}nez$^{a,}$\footnote{E-mail address: IFFJIMENEZ
@ROCA.CSIC.ES.} and Germ{\'a}n Sierra$^{b,}$\footnote{E-mail address: SIERRA
@CC.CSIC.ES.}}
\vspace{0.5 true  cm}

\normalsize
$^{a}${\em Departamento de Aerotecnia,\\
  Universidad Polit{\'e}cnica de Madrid. 28040-Madrid, Spain.}\\
$^{b}${\em Instituto de Matem{\'a}ticas y F{\'\i}sica Fundamental,}\\
{\em  Consejo Superior Investigaciones Cient{\'\i}ficas. 28006-Madrid, Spain.}

\end{center}
\vspace{1 true  cm}

\begin{abstract}

Fokker-Planck equations have been applied in the past to field theory topics
such as the stochastic quantization and the stabilization of bottomless action
theories. In this paper we give another application of the FP-techniques in a
way appropriate to the study of
the ground state, the excited states and the critical behaviour of quantum
lattice Hamiltonians.
With this purpose, we start by considering a discrete or lattice version of
the standard FP-Hamiltonian. The well known exponential ansatz for the ground
state wave functional becomes in our case an exponential ``cluster'' expansion.
With a convenient choice for this latter, we are able to construct
FP-Hamiltonians
which to a large extent reproduce critical properties of ``realistic'' quantum
lattice
Hamiltonians, as the one of the Ising model in a transverse field (ITF).
In one dimension, this statement is made manifest by proving that the
FP-Hamiltonian
we built up belongs to the same universality class as the standard
ITF model or, equivalently, the 2D-classical Ising model. To this
respect, some
considerations concerning higher dimensional ITF models are outlined.

\vspace{0.8cm}

\end{abstract}

\end{titlepage}

\newpage

\section{Introduction}

A fundamental problem in Statistical Mechanics and Solid State Physics
is the search of the ground state and excitations of many body Hamiltonians.
Exact solutions have  been obtained for integrable one dimensional Hamiltonians
\cite{Baxter},\cite{Lieb}, \cite{Pfeuty1}. However, the majority of the models
are not integrable, specially
in dimensions greater than one, and in these cases one has to resort
to approximation techniques.
We have squematically,

\ho
H \,\,\,({\rm Hamiltonian}) \longrightarrow \left\{\begin{array}{l}
{\rm Perturbation\,\, theory}\\
{\rm Variational\,\, methods}\\
{\rm Renormalization \,\,group}\\
{\rm Numerical\,\, approaches}\\
\hspace{2.1cm}\vdots
\end{array}
\right \}\longrightarrow \mid \psi_{0} \rangle\,\,\, ({\rm ground\,\,\,
state})
\label{0.1}
\la

The path from $H$ to $\mid \psi_{0} \rangle$ is very cumbersome
and can only be fully accomplished in a few cases. Instead, one could
try to reverse
the arrows in (\ref{0.1}),  by  making  a clever choice of the state
 $\mid \psi_{0} \rangle$ first,
and asking latter for the Hamiltonian $H$ for which the proposed
ansatz becomes the exact ground state (this second strategy can be
paraphrased in the words of
\cite{Arovas} as ``finding questions to some interesting answers''. Some
remarkable cases where this
``inverse method'' has been applied would include the exponential Jastrow
wave functions, used in the study of liquid $^{4}$He \cite{Mahan} and nuclear
matter \cite{Jastrow}, the valence-bond solid states (VBS) \cite{Affleck},
the Laughlin wave function
\cite{Lau}, etc. The path from $\mid \psi_{0} \rangle$ to $H$ could be
summarized as
follows:

\ho
\mid \psi_{0} \rangle \longrightarrow
\left\{\begin{array}{l}
R_{\alpha} \mid \psi_{0} \rangle = 0\\
\alpha = 1,\cdots,N
\end{array}
\right \}\longrightarrow
\left\{\begin{array}{l}
H= \sum_{\alpha =1 }^{N} R_{\alpha}^{\dagger} R_{\alpha}\\
H \mid \psi_{0} \rangle = 0
\end{array}
\right \}
\label{0.2}
\la

\noindent where $\alpha$ denotes either sites in a lattice, or
particles or
some other quantum numbers. In words, one has first to
find a set of operators $R_{\alpha}$ which annihilate
the state $\mid \psi_{0} \rangle$; afterwards one builts up a
positive semidefinite Hamiltonian $H$  according to the prescription
(\ref{0.2}). If the
state  $\mid \psi_{0} \rangle$ of (\ref{0.2}) is a valence-bond state,
then the $R_{\alpha}$ are projector operators (i.e. $R_{\alpha}^{\dagger} =
R_{\alpha}$, $R_{\alpha}^{2}= R_{\alpha}$) and we can write
$H=\sum_{\alpha}R_{\alpha}$.
However, for Jastrow type wave functions,
the $R_{\alpha}$ are not projectors, and these will be the kind of operators
which we shall be concerned with.

The aim of this paper is to combine the direct and inverse methods
(\ref{0.1})-(\ref{0.2}) in the study of quantum lattice
Hamiltonians. To be more concrete, the problem we want to
adress is the construction of the ground state $\mid \psi_{0} \rangle$
and the excited states of a quantum Hamiltonian $H$ defined on a
lattice. The strategy
could be summarized in the following diagram:

\ho
H \stackrel{\begin{array}{l}\rm Perturbation\\ \,\,\,\,\,\,\, \rm theory
\end{array}}{\longrightarrow}
\psi \,\,\,({\rm Exponential \,\,\,ansatz})
\stackrel{\begin{array}{l}\rm Fokker-Planck\\\,\,\,\,\,\,\,\, \rm method
\end{array}}{\longrightarrow}H^{FP}
\label{0.3}
\la

The first step consists in the construction of a state of exponential
type, which will be our candidate for the
ground state of $H$. To make this ansatz, we shall follow
 the perturbative-variational procedure of determination of
 the ground state $\mid \psi_{0}(\lambda) \rangle$ of a lattice Hamiltonian,
in the form
$H(\lambda)=H_{0}+ \lambda H_{1}$,
established in \cite{German}; here $H_{0}$ is a Hamiltonian
whose non degenerated ground state $\mid 0\rangle$ is known,
and $\lambda$ is a coupling constant.

Two alternative ways for writting the state $\mid \psi_{0}(\lambda)
\rangle$
of $H(\lambda)$
were proposed in reference  \cite{German}, to wit:

\ho
\mid \psi_{0} (\lambda) \rangle = \exp (\sum_{n\geq 1} \lambda^{n}
U_{n}) \mid 0\rangle
\label{0.4}
\la

\noindent  or

\ho
\mid \psi_{0} (\lambda) \rangle = \exp(\sum_{I} \alpha_{I} (\lambda)V_{I})
 \mid 0\rangle
\label{0.5}
\la

\noindent where both $U_{n}$ and $V_{I}$ are a certain set of
interrelated local operators (see below), whereas the $\alpha_{I}$
 are certain functions of $\lambda$.

Expansion (\ref{0.4}) is adequate to make contact with the standard
perturbation theory (PT), whereas expansion (\ref{0.5}) is a kind of
cluster expansion.
Let us describe in some more detail the main ideas of \cite{German},
since this paper is,
after all, an extension of that work.

 When one expands the exponential of (\ref{0.4})
and compares the result with the usual PT,
one obtains a set of equations for the operators
$U_{n}$. Thus, we can say that PT to order
$\nu$  fixes , up to some freedom, the familly of operators
$\{ U_{n}\}_{1\leq n \leq \nu}$. Working through particular examples,
as the Ising model in a transverse field, one observes that each
operator $U_{n}$
can in fact be written as a  sum of ``irreducible'' operators $V_{I}$:

\ho
U_{n}= \sum_{I} p_{n,I} V_{I}
\label{0.6}
\la

\noindent where in \cite{German} the label $I$ denoted  a cluster formed
by a finite number of
1's distributed on certain sites of the lattice, the rest of sites
being occupied by zeros.
The operator $V_{I}$ obeys a fusion algebra given by $V_{I} V_{J} =
V_{I+J}$, in such a way
that the cluster $I+J$ is obtained through the usual ${\bf Z}_{2}$-sum rules:
$0+0 = 1 + 1 = 0$\,\,\,, 0 + 1 = 1 + 0 = 1 (this rule has its origin
in the ${\bf Z}_{2}$-symmetry
of the Ising model). The usefulness of eq.(\ref{0.6}) lies in the fact
that a given cluster operator
$V_{I}$ may appear at
different -and possible at all- orders in PT. This would imply that
for a fixed value of $I$,\,\,
$p_{n,I}$ might have an infinite number of non zero values. Introducing
(\ref{0.6}) in (\ref{0.4}) we can relate the perturbative  and
cluster expansions with:

\ho
\alpha_{I}(\lambda) = \sum_{n\geq \nu_{I}} p_{n,I} \lambda^{n}
\label{0.7}
\la

\noindent where the number

\ho
\nu_{I}= {\rm minimun} \{n / p_{n,I} \neq 0\}
\label{0.7bis}
\la

\noindent is called  the level of the operator $V_{I}$; somehow
$\nu_{I}$ measures the size
of the cluster \cite{German}.
Definition (\ref{0.7}) implies that $\alpha_{I}(\lambda)$ has a
Taylor expansion starting at $\lambda^{\nu_{I}}$ and containing
non vanishing terms of higher powers whenever $V_{I}$ appears in $U_{n}$.

If we think of $p_{n,I}$ as a matrix with two entries $n$ and
$I$ -or rather $\nu_{I}$-, we see by its definition that $p_{n,I}$ is
a lower triangular matrix of infinite dimension (see fig.1a).
Thus PT to order $\nu$ gives
all the entries of $p_{n,I}$ to
 a ``depth'' $\nu$, i.e $\{ p_{n,I}\}_{1\leq n \leq \nu}$ (see fig.1b).

 However, eq.(\ref{0.5})
provides another approach to the exact ground state. Indeed,
the knowledge of a single function $\alpha_{I}(\lambda)$
for a given operator $V_{I}$ involves all orders in PT.
This result suggests a non-perturbative approach to
$\mid \psi_{0} \rangle$ based on the truncation of (\ref{0.5})
to just those operators $V_{I}$ with an order less or equal
to a given order $\nu$ (see fig.1c):

\ho
\mid \psi^{(\nu)}\rangle  = \exp(\sum_{I,\,\,\nu_{I}\leq \nu}
\alpha_{I}^{(\nu)} V_{I})
\mid 0\rangle
\label{0.8}
\la

In \cite{German} the functions $\alpha_{I}^{(\nu)}$ were fixed
by using the state $\mid \psi^{(\nu)}\rangle$ as a trial
variational state for the Hamiltonian $H(\lambda)=H_{0}+ \lambda H_{1}$.
In this case the variational ansatz agrees to order $\nu$ with the
perturbative ground
state and gives non perturbative predictions for large values of
$\lambda$, where
PT cannot be trusted.

In this paper we propose an alternative way for computing
the parameters $\alpha_{I}^{(\nu)}$.
The principle of our method is a generalization of the well known
Fokker-Planck's procedure, by means of which one associates a Hamiltonian
to a wave function having an exponential form (see appendix A, for
an application to Quantum Mechanics). The rule is simply
to act with $H_{0}$ on the proposal (\ref{0.8}) until
 one arrives to a Fokker-Planck Hamiltonian
$H^{FP,(\nu)}(\alpha_{I}^{(\nu)})$ which depends, of course,
on the whole set of parameters $\alpha_{I}^{(\nu)}$
entering in the definition of $\mid \psi^{(\nu)}\rangle$.
The condition we use to fix these parameters as functions of $\lambda$ is that
$H^{FP,(\nu)}$ should only differ from $H$
in operators of a level greater than $\nu$, i.e:

\ho
H-H^{FP,(\nu)}= \sum_{I,\nu_{I}\geq \nu}
 C_{I}^{(\nu)}(\lambda) V_{I}
\label{0.9}
\la

Correspondingly, the FP-energy $E^{FP,(\nu)}(\lambda)$
although incorporating non perturbative effects,
should agree
with the exact ground state energy up to order $\lambda^{\nu +1}$.
 Compairing $E^{FP,\nu} (\lambda)$ (for $\nu = 1$ and $2$) with
 the variational energy $E^{var,\nu} (\lambda)$ obtained in \cite{German}
in the study of the Ising model in a transverse field, one observes
that the latter gives better values than the former, as to the energy
is concerned.
It seems natural then to
wonder about the advantages of the FP-approach versus the variational one.
Well, the problem is that within the framework of
the variational approach is very difficult to
go to higher values of $\nu$ since, for example, the norm of the variational
state (\ref{0.8}) turns out to be the partition function of a
statistical mechanical model whose couplings are dictated by the
operators $V_{I}$.
On the contrary, in the FP procedure the norm of the state (\ref{0.8})
is not needed to
fix the parameters $\alpha_{I}^{(\nu)}$ in terms of $\lambda$; rather,
what is required
is the algebra satisfied by the cluster operators $V_{I}$, and this algebra
 is relatively simple in the example considered in this paper. Besides, within
the FP-method we can always match the perturbative results to any order,
at the expense of including a sufficient number of operators in the ansatz
(\ref{0.8}). It is clear, however, that except for the lowest orders
in PT the complexity
of the ansatz will obscure its usefulness. On the other hand, one
of the aims in
constructing more general ansatzs for ground states is to see whether
or not they exhibit
critical behaviour. In this sense, what actually matters is to determine the
universality class which this Hamiltonians belongs to, rather than their
apparent simplicity. Bearing in mind these ideas, we can try to reinterpret
eq.(\ref{0.3}) as a map from a given Hamiltonian $H$ to
a FP-Hamiltonian $H^{FP}$ belonging to the same universality class as $H$;
in this case $H$ and $H^{FP}$ would share the same critical properties.
To achieve this map one should consider ansatzs involving cluster operators
$V_{I}$ with all the possible sizes (see fig.1d). We shall show in this paper
an example where such a construction can be accomplished.
\vspace{0.5cm}

The organization of the paper is as follows. In section 2 we review
the application
of the FP machinery to field theory in the continuum
and reconsider it in such a way that it
can be applied to lattice Hamiltonians.
In section 3 we benefit from this generalization to look into the Ising
model in a transverse
field (ITF) in  $D=1$ and $D=2$ dimensions. In section 4 we generalize
the exponential
ansatz of section 3 in order to include operators at all length scales,
and we show in the one
dimensional case
the existence of a critical behaviour for a particular value of the
coupling constant.
In section 5 we study the problem of the excited states of FP-Hamiltonians
using a ``change of variables'' which brings the Hamiltonian to a simpler form
$H^{RFP}$,  which we call ``reduced'' FP-Hamiltonian. Finally,
a general discussion with  prospects for the future is proposed in section 6.

 We have also included four appendices
which illustrate in detail some aspects mentioned in the course of the paper.
In particular we collect in appendix D the results of sections 3 and 4
extended to the case of ITF models in dimensions higher than one.

\setcounter{equation}{0}
\section{ Fokker-Planck lattice hamiltonians:
\newline General construction}

The Fokker Planck Hamiltonians arise naturally in the study of
stochastics processes in Statistical Mechanics or
Quantum Field Theories. The idea is to consider a particle or a field
evolving in time under the combined action of a force and a gaussian noise
\cite{Paris}. The evolution
equation for such  a field $\phi (x,t)$ is given by the Langevin equation:

\ho
\frac{\partial \,\phi (x,t)}{\partial t} = - \frac{\delta \,S[\phi]}
{\delta \phi}
+ \eta (x,t)
\label{p1}
\la

\noindent where $S[\phi]$ is the potential and $\eta (x,t)$ is the noise.
Here $x\in {\bf R}^{D}$ is the space location of the field. The corresponding
probability $P[\phi,t]$ for finding  a given configuration
$\phi (x)$ at the time $t$ is governed by the Fokker-Planck (FP)
equation. A suggestive way of writing this equation comes up
if we make the change of variables
$P[\phi,t] = e^{-\frac{S[\phi]}{2}}\,\, \psi[\phi,t]$; with this change,
the FP equation
for $\psi[\phi,t]$ becomes a Schr{\"o}dinger like equation in imaginary
time, namely:

\ho
\frac{\partial \psi[\phi,t]}{\partial t} = - H^{FP} \psi[\phi,t]
\label{p2}
\la

\noindent where

\ho
H^{FP}=\frac{1}{2}\int d^{D}x [- \frac{\delta^{2}}{\delta {\phi}^2}
+\frac{1}{4}
(\frac{\delta S}{\delta \phi})^2 -\frac{1}{2}\frac{{\delta} ^2 S}
{\delta {\phi}^2}]
\label{3}
\la

\noindent is called the Fokker-Planck Hamiltonian. To this point,
several comments are in order:

\begin{description}

\item {(i)} $H^{FP}$ is a positive semidefinite operator. This follows
inmmediately if we
write the Hamiltonian $H^{FP}$ as :

\ho
H^{FP}=\frac{1}{2}\int d^{D}x R^{\dagger}(x) R(x)
\label{4}
\la

\noindent where

\ho
R(x)=i\frac{\delta}{\delta \phi(x)} + \frac{i}{2}\frac{\delta S}{\delta
\phi(x)}
\label{5}
\la

\item{(ii)} The functional $\psi_{0}[\phi] = e^{- \frac{S[\phi]}{2}}$ is
annihilated by the family of commuting operators $R(x)$ (i.e. $R(x)
R(y) = R(y) R(x)$ )

\ho
R(x)  \psi_{0}  = 0, \,\,\,\,\,\,\,\,\forall x
\label{pp4}
\la

Hence, if $\psi_{0}$ is a normalized state, then it becomes
a ground state
of the FP-Hamiltonian (\ref{3}).

\item{(iii)} The norm of the ground state $\psi_{0}[\phi]$ can be
interpreted as
the partition function of a statistical mechanical model or euclidean quantum
field theory in $D$-dimensions, with Boltzmann weights $e^{-S[\phi]}$

\ho
Z\equiv \langle \psi_{0}\mid \psi_{0}\rangle=\int D\phi \,\, e^{-S[\phi]}
\label{1}
\la

\end{description}

These simple properties have been used in a variety of applications,
some of which are worth
mentioning. To begin with, let us quote the stochastic quantization
method of Parisi and Wu
\cite{Parisi}. Introducing a ficticious time (the fith-time), this procedure
enables one to compute quantum mechanical expectation values of
observables $Q[\phi]$
in terms of time averages of fields evolving under the Langevin
equation (\ref{p1}):

\ho
\langle Q\rangle = \frac{1}{Z} \int D\phi \,\, Q[\phi]\,\,e^{-S[\phi]} =
\lim_{T \rightarrow \infty} \frac{1}{T} \int_{0}^{T} Q[\phi(x,t)] d\,t
\label{p4}
\la

This formulation has been applied to quantum field theories,
including gauge
theories.

Another fruitful application carried out in reference \cite{Halpern}
concerns to the stabilization of bottomless field theories, i.e. theories
with a potential unbounded from below and, therefore, with divergent
partition functions. Here one uses the unbounded action $S[\phi]$
and eq.(\ref{3}) to
construct a FP-Hamiltonian, which thanks to property (i) is well defined.
Since $H^{FP}$ is bounded from below, the corresponding ground state
$\psi_{0}^{nor}$
will be normalizable; that gives rise to a bounded  effective action
$S_{eff}[\phi]$ through
the definition  $\psi_{0}^{nor}\equiv  e^{-\frac{S_{eff}[\phi]}{2}}$.
This stabilization
procedure has also been applied to a number of theories including gravity
\cite{Green1},  and $2D$-matrix models
\cite{Mari}, \cite{Miramon}

Finally, we want to mention the connection between FP-Hamiltonians and
supersymmetry. Let us suppose that we enlarge the field theory
presented earlier with a pair
of anticommuting fields $\theta(x)$ and $\bar{\theta}(y)$ satisfying
the anticommutation
relations:

\ho
\{\theta(x), \theta(y)\} = \{\bar{\theta}(x), \bar{\theta}(y)\} = 0
\label{pp5}
\la

\ho
\{\theta(x), \bar{\theta}(y)\} = \delta (x - y)
\label{pp6}
\la

Now we define two supersymmetric charges $Q$ and $\bar{Q}$

\ho
Q= \int dx \,\,R(x) \theta(x)
\label{pp7}
\la

\ho
\bar{Q} \equiv Q^{\dagger} = \int dx \,\, R^{\dagger} (x) \bar{\theta} (x)
\label{pp8}
\la

\noindent and satisfying

\ho
Q^{2} = {\bar{Q}}^{2} = 0
\label{ppp8}
\la

Then it is easy to see that the eq.(\ref{pp4})  characterizing the
state $\psi_{0}$
can be translated into the equations

\ho
Q \psi_{0} = \bar{Q} \psi_{0} = 0
\label{ppp9}
\la

In terms of $Q$ and $\bar{Q}$ one can define a supersymmetric
Hamiltonian $H_{SS}$
whose restriction to the bosonic sector
(i.e $\bar{\theta} = 0)$ is nothing else but the FP-Hamiltonian
(\ref{3}), namely:

\[H_{SS} = \frac{1}{2} (Q \bar{Q} + \bar{Q} Q) \]
\ho
= H^{FP} + \frac{1}{2} \int d^{D} \,x\,\, \frac{\delta^{2} S}{\delta
\phi(x)^{2}}
[\theta(x), \bar{\theta}(x)]
\label{pp9}
\la

In the SUSY-terminology we can say that the exponential state $\psi_{0} =
e^{-\frac{S[\phi]}{2}}$ is a supersymmetric ground state \cite{Witten}.

The connection between the FP-method and SUSY is very suggestive but
we shall not
pursued it here, except for a discrete version of eqs.(\ref{pp5})-(\ref{pp9}).
 \vspace{0.5cm}

As for the use of FP-ideas we make in this paper, let us say that they
can be simply illustrated in the continuum as follows. Suppose we are given a
field theoretical hamiltonian:

\ho
H=\int d^{D}x [- \frac{1}{2}\,\,\frac{\delta^{2}}{\delta {\phi}^2} +U(\phi)]
\label {6}
\la

\noindent where $U(\phi)$ is a funtion of $\phi(x)$, $\nabla \phi(x)$, etc.
Then if we look for the ground state of $H$ in the form
$\psi_{0}[\phi] = e^{- \frac{S[\phi]}{2}}$ ,
it turns out that the potential $U(\phi)$ and the ground state energy
$E_{0}$ are related
to $S(\phi)$ by the eq.

\ho
U(\phi) - E_{0}= \frac{1}{8} (\frac{\delta S}{\delta \phi})^2
-\frac{1}{4}\frac{{\delta} ^2 S}{\delta {\phi}^2}
\label{7}
\la

Eq. (\ref{7}) is a functional version of the Ricatti equation
\cite{Turbiner}. Consequently, the
eigenvalue problem for the ground state of $H$ is reduced, assuming the
exponential ansatz
$\psi_{0}[\phi] = e^{- \frac{S[\phi]}{2}}$,
to the solution of the functional Ricatti equation (\ref{7}).

In the context of Quantum Mechanics the analogue of eq.(\ref{7}) can be
taken as the starting
point of a perturbative-variational method \cite{Turbiner} very close in
spirit to the one
developed in \cite{German}. There are situations where the field
theoretical Hamiltonian
(\ref{6}) can be diagonalized exactly, for example in terms of creation
and annihilation
operators as in pure QED \cite{Greensi}. The corresponding wave
functional $\psi_{0}[\phi]$
is  then essentially a gaussian. When interactions are present,
matters are not so simple. In any case, before looking for the ground state
of quantum field Hamiltonians defined in the continuum,
 one should first regularize the model so as to control
the infinities and this can be done, for example, by using a lattice
regularization.
This is, therefore, another motivation to consider the FP-method
in the context of quantum lattice Hamiltonians.

\vspace{0.8cm}

In the remaining of this section we shall present a generalization of the FP
techniques introduced above. From an abstract point of view, the
basic elements
required in a FP-construction are: a) an algebra of operators, b) a vacuum
state $\mid 0 \rangle$, c) a free Hamiltonian $H_{0}$ and d) an action $S$.
In the continuum FP-construction, these elements are given respectively
by the operators $\frac{\delta}{\delta \phi(x)}$ and $\phi(x)$, the free
Hamiltonian $H_{0}= -\frac{1}{2}\int d^{D} x \,\,\frac{\delta^{2}}
{(\delta \phi(x))^{2}}$
whose ground state is the constant wave function, and the action $S[\phi]$
which  is a functional of the field $\phi(x)$.
\vspace{0.3cm}

The construction can be implemented as follows:

\vspace{0.8cm}
{\bf a) Algebra of operators.} We consider a family of unitary
operators $\{ X_{a}\}_{a \in {\cal C}_{0}}$, $\{ Z_{A}\}_{A \in {\cal C}_{1}}$
satisfying the relations:

\[X_{a} X_{b} = X_{b} X_{a}\]
\ho
Z_{A} Z_{B} = Z_{B} Z_{A}
\label{9}
\la
\[X_{a} Z_{A} = \omega_{aA} Z_{A}X_{a}\]

\noindent where the $\omega_{aA}$ are phases ($\mid \omega_{aA} \mid = 1$).
The labels $a$ and $A$ run, respectively,
over discrete sets ${\cal C}_{0}$ and  ${\cal C}_{1}$, whose nature
 will be specific for each model. In practical examples they will denote sites,
links, plaquettes or internal quantum numbers.

We shall demand that if ${\cal C}_{0}$ and  ${\cal C}_{1}$
contain an operator, say $X_{a}$ or $Z_{A}$, then
they must also contain their hermitean conjugated, i.e:

\[X_{a}^{\dagger} = X_{\bar{a}}\,\,\,\,\,\,,\,\,\,\,\,\, a, \bar{a} \in
{\cal C}_{0}\]
\ho
Z_{A}^{\dagger}= Z_{\bar{A}}\,\,\,\,\,\,,\,\,\,\,\,\, A, \bar{A} \in
{\cal C}_{1}
\label{10}
\la

The Hilbert space where the operators $X_{a}$ and $Z_{A}$ are acting
will be denoted by ${\cal H}$. The algebra (\ref{9}) is actually the algebra
satisfied by the order-disorder variables of, for example, the Ising model,
the Potts model or the ${\bf Z}_{N}$ gauge models.

\vspace{0.8cm}
 {\bf b) Vacuum vector.} We shall suppose that in the Hilbert space ${\cal H}$
there is a  vacuum vector $\mid 0 \rangle$ satisfying:

\ho
X_{a} \mid 0 \rangle = \mid 0 \rangle \,\,\,\,\,\,,\,\,\,\,\,\, \forall
a\in {\cal C}_{0}
\label{11}
\la

The existence of a common eigenvector for all the $X_{a}$-operators is
in principle
possible, because all these operators commute among themselves.
Nevertheless, this vector
needs not to be unique, i.e. there may exist other vectors satisfying
(\ref{11}); this will
be the case of models with a broken phase.

\vspace{0.8cm}
  {\bf c) Free Hamiltonian $H_{0}$.} Let us define the following positive
semidefinite Hamiltonian:

\ho
H_{0}= \frac{1}{2} \sum_{a \in {\cal C}_{0}} (X_{a}^{\dagger} -
{\bf 1})\,\, (X_{a} - {\bf 1})
\label{12}
\la

\noindent which, using the unitarity properties of the $X_{a}$
operators and eq.(\ref{10})
transforms in

\ho
H_{0}=  \sum_{a \in {\cal C}_{0}} ({\bf 1} - X_{a} )
\label{13}
\la

With this definition, it is clear that the vacuum state $\mid 0 \rangle$
is a ground
state of $H_{0}$

\ho
H_{0} \mid 0 \rangle = 0
\label{14}
\la

\vspace{0.8cm}
{\bf d) Exponential ansatz.} Suppose we are given a generic operator $S(Z_{1},
Z_{2},\cdots) = S[Z_{A}]$, which we shall call the action operator, and
 constructed out of the family of commuting
$Z$-operators. From $S[Z_{A}]$ we
introduce the state:

\ho
\mid \psi_{0} \rangle = e^{- \frac{1}{2} S[Z_{A}]} \mid 0 \rangle
\label{15}
\la

It is easy to check that the action of an operator $X_{a}$ on $\mid 0
\rangle$ is given by:

\[X_{a} \mid \psi_{0} \rangle = e^{-\frac{1}{2} S[\omega_{aA}
Z_{A}]} X_{a} \mid 0 \rangle \]
\ho
= e^{-\frac{1}{2} S[\omega_{aA}
Z_{A}]}\,\, e^{\frac{1}{2} S[Z_{A}]}\mid \psi_{0} \rangle \equiv
e^{\Delta_{a}(S)} \mid \psi_{0} \rangle
\label{16}
\la

\noindent where we have defined

\ho
\Delta_{a}(S) = \frac{1}{2}(S[Z_{A}] - S[\omega_{aA} Z_{A}])
\label{17}
\la

For latter purposes we shall write eq. (\ref{16}) as

\[R_{a} \mid \psi_{0} \rangle = 0\]
\ho
R_{a} \equiv e^{- \Delta_{a}(S)}\,X_{a} - {\bf 1}
\label{18}
\la

Incidentally, it is noteworthy the commutation relation for the $R$ operators

\ho
[R_{a},R_{b}] = 0
\label{18m}
\la

At this moment we may wonder whether there is a Hamiltonian
for which the state $\mid \psi_{0} \rangle$ becomes an exact ground state.
There are several answers to this question but in this paper we
propose the following:

\vspace{0.8cm}
{ \bf Fokker-Planck Hamiltonian}. Let $\Omega_{a}$ be a hermitian
($\Omega_{a}^{\dagger} =
\Omega_{a}$) and positive semidefinite operator ($\Omega_{a} \geq 0$); then
we shall call a FP-Hamiltonian the following hermitian positive
semidefinite operator:

\ho
H^{FP} = \frac{1}{2} \sum_{a \in {\cal C}_{0}} R_{a}^{\dagger} \Omega_{a}
R_{a}
\label{19}
\la

Obviously the state (\ref{15}) is a ground state for this Hamiltonian

\ho
H^{FP} \mid \psi_{0} \rangle = 0
\label{20}
\la

Unfortunately, the choice $\Omega_{a}=1$  although natural, gives rise
to Hamiltonians where the
operators $X_{a}$ and $Z_{A}$ are mixed together. Yet, a peculiarity of
this choice is that the corresponding FP-Hamiltonian presents a
hidden supersymmetry.
Following the same steps as in the continuum, this suppersymmetry can
be made manifest
if we introduce a discrete set of anticommuting operators $\theta_{a}$ and
${\bar{\theta}}_{b}$ satisfying:

\ho
\{\theta_{a}, \theta_{b}\} = \{{\bar{\theta}}_{a},{\bar{\theta}}_{b}\} = 0
\label{20p}
\la

\ho
\{\theta_{a}, {\bar{\theta}}_{b}\} = \delta_{ab}
\label{21p}
\la

The supersymmetric generators analogues to eqs.(\ref{pp7})-(\ref{pp8}) read now

\ho
Q = \sum_{a \in {\cal C}_{0}} R_{a} \theta_{a}
\label{22p}
\la

\ho
\bar{Q} = \sum_{a \in {\cal C}_{0}} R^{\dagger}_{a} {\bar{\theta}}_{a}
\label{23p}
\la

\noindent and verify $Q^{2} = {\bar{Q}}^{2} = 0$.

The SUSY FP-Hamiltonian is then

\[
H^{FP}_{SS} = \frac{1}{2} \,\,(Q \bar{Q} + \bar{Q} Q)\]
\ho
= \frac{1}{2}\sum_{a \in {\cal C}_{0}} R^{\dagger}_{a} R_{A}
+ \frac{1}{2} \sum_{a \in {\cal C}_{0}} [R_{a}, R^{\dagger}_{a}] \,\,
\theta_{a} {\bar{\theta}}_{a}
\label{24p}
\la

Hence in the bosonic sector $H^{FP}_{SS}$ reduces to eq.(\ref{19}) with
$\Omega_{a} = 1$.
The problem with the Hamiltonian (\ref{24p}) is that
it mixes the $X$ and $Z$ operators in a complicated form, namely:

\ho
H^{FP}_{\Omega_{a} = 1} = \sum_{a \in {\cal C}_{0}}
 \cosh ( \Delta_{a}(S) )\,\,(e^{\Delta_{a}(S)} - X_{a})
\label{25p}
\la

For this reason, and in spite of its intrinsic theoretical interest,
we shall not
pursue the study of (\ref{24p}) in this paper.

{}From now on,  we shall consider the choice $\Omega_{a}= e^{\Delta_{a}(S)}$,
which is an hermitian and positive semidefinite operator whenever
$S$ is hermitian.
In this case, $H^{FP}$ takes a simple form

\ho
\left\{\begin{array}{l}
\Omega_{a}= e^{\Delta_{a}(S)}\\
S^{\dagger}= S
\end{array}\right \}
\,\,\,\,\,\,\longrightarrow\,\,\,\,\,\,H^{FP}=
\sum_{a \in {\cal C}_{0}} (e^{\Delta_{a}(S)} - X_{a})
\label{21}
\la

\noindent or in words, $H^{FP}$ is the sum of the free Hamiltonian
(\ref{13}) and
an interacting Hamiltonian which depends only on the $Z$-operators.

\vspace{0.5cm}
The previous results are the starting point of  our method.
Thus, let us suppose we are given a Hamiltonian of the type

\ho
H = - \sum_{a \in {\cal C}_{0}} X_{a} + U[Z_{A}]
\label{22}
\la

\noindent and whose ground state $\psi_{0}$ we would like to determine
in the usual way:

\ho
H \psi_{0} = E_{0} \psi_{0}
\label{23}
\la

A solution of this problem arises if one is able to find an action
operator $S[Z_{A}]$ which satisfies the Ricatti equation

\ho
U[Z_{A}] - E_{0} = \sum_{a \in {\cal C}_{0}} e^{\Delta_{a}(S)}
\label{24}
\la

For the eigenvalue problem (\ref{23}) the operator $U$ as
defined in eq.(\ref{24}) is a datum whereas $S$ is an unknown;
however, as it happens in the continuum, one can reverse
the roles and take eq.(\ref{24}) as defining the interaction
term which corresponds to a given state of the form (\ref{15}).
In the sequel, we shall use both kinds of interpretations.

Finally, to complete our generalization of the FP-method
we shall study the norm of the exponential state (\ref{15}).
If $S$ is an hermitian operator, the norm of this state
can be considered as the partition function
of an statistical mechanical model with real and positive
Bolztmann weights. To show this, we can select an orthonormal
basis $\{\mid I \rangle\}$ of the Hilbert space $\cal H$
where the $Z$-operators are diagonal

\ho
Z_{A} \mid I \rangle = z_{A,I} \mid I \rangle
\label{25}
\la

Then the norm of the state (\ref{15}) becomes:

\ho
Z = \langle \psi_{0} \mid \psi_{0} \rangle =
\langle 0 \mid e^{- S[Z_{A}]} \mid 0 \rangle =
\sum_{I} \mid \langle 0 \mid I \rangle {\mid}^{2}e^{- S[z_{A,I}]}
\label{26}
\la

In most of the examples $ \langle 0 \mid I \rangle$ will be a constant
and that implies that the Boltzmann weights are given by
$e^{- S[z_{A,I}]}$.

\vspace{0.8cm}
Before we apply this construction to particular examples,
let us mention that a
 realization of the algebra (\ref{9}) in the continuum can be developed
in terms of the Weyl operators:

\[
X_{\vec r} = e^{i \hbar \Pi (\vec r)} = e^{\hbar
 \frac{\delta}{\delta \phi(\vec r)}}\]
\ho
Z_{\vec r} = e^{\phi (\vec r)}
\label{27}
\la

\noindent These operators satisfy the algebra

\ho
X_{\vec r_{1}} Z_{\vec r_{2}}= e^{i \hbar \delta(\vec r_{1}
- \vec r_{2})} Z_{\vec r_{2}} X_{\vec r_{1}}
\label{28}
\la

\noindent but obviously this expression should
be regularized in order to build a consistent theory.

The FP Hamiltonian given by eq.(\ref{3})
 can be obtained just by taking the limit $\hbar \rightarrow 0$
of the corresponding (\ref{21}) (see Appendix A).
Thus our generalization accounts to a ``deformation''
 of the usual FP-approaches. The word deformation is used
here with a meaning very similar to the one used in the theory
of quantum groups \cite{Drin}, \cite{Jimbo}.

 \setcounter{equation}{0}
\section{ The Ising model in a transverse field }

As a prior example of the usefulness of the FP-method outlined in the
previous section, let us consider the case of the $D$-dimensional
Ising model in a transverse field (ITF).

The model is defined on a hypercubic lattice ${\bf Z}_{L}^{D}$ with
$N=L^{D}$ sites and Hamiltonian \cite{Gennes},\cite{Pfeuty2},\cite{Stich}:

\[H_{Ising,D}= H_{0} + U_{Ising,D}\]
\ho
=-\sum_{{\bf j}} \sigma_{{\bf j}}^{x} - \lambda \sum_{{\bf j},\mu}
\sigma_{{\bf j}}^{z} \sigma_{{\bf j}+\mu}^{z}
\label{33}
\la

\noindent where $\sigma_{{\bf j}}^{x}$, $\sigma_{{\bf j}}^{z}$ are
Pauli matrices
acting at the site ${\bf j}$, with ${\bf j}= (j_{1},\cdots, j_{D})$,
 $j_{a}= 1,\cdots,L$
and $\mu$ is any of the lattice vectors $\mu_{1}=(1,0,\cdots,0),\cdots,
\mu_{D}=(0,0,\cdots,1)$. These matrices give a representation of
the algebra (\ref{9}) for $k=2$. We shall assume in (\ref{33})
periodic boundary
conditions, i.e $\sigma_{{\bf j}+ L\mu_{i}}^{x,z}=
\sigma_{{\bf j}}^{x,z},\,\,\,\,\forall i$.

It is well known that the ground state of (\ref{33}) for $\lambda$ smaller
than a certain $D$-depending critical value $\lambda_{c}$ is non degenerate.
This accounts to a disordered regime where the order parameter
$\langle \sigma_{{\bf j}}^{z}\rangle $ vanishes.
For $\lambda > \lambda_{c}$, however, the ${\bf Z}_{2}$-symmetry is broken
spontaneously and there are two ground states in the termodynamic limit with
$\langle \sigma_{{\bf j}}^{z}\rangle \neq 0$. This is the ordered
regime of the ITF model.

Let us focus our discussion on the disordered region, where the
ground state is non degenerate. Our aim is, for the moment, to construct
this ground state.

It is easy to check that all the ingredients needed to apply the
generalized FP-construction
are fulfilled by the ITF model. Indeed the algebra of operators $X$ and $Z$ are
given by the Pauli matrices $\{\sigma_{j}^{x}\}$ and
$\{\sigma_{j}^{z}\}$, respectively.
The vacuum vector $\mid 0 \rangle$ is defined by:

\ho
\mid 0 \rangle = \prod_{j=1}^{N} \mid 0 {\rangle}_{j}
\label{34}
\la

\noindent where

\ho
\sigma_{j}^{x} \mid 0  {\rangle}_{j}= \mid 0 {\rangle}_{j}
 ,\,\,\,\forall j=1,\cdots,N
\label{34bis}
\la

The free Hamiltonian (\ref{13}) is, apart from a constant, the same as
the free Hamiltonian
in eq.(\ref{33}). The problem is therefore to look for the ground state
of (\ref{33})
in terms of the exponential ansatz

\ho
\mid \psi_{0} \rangle = e^{-\frac{1}{2} S(\sigma_{1}^{z},\cdots,
\sigma_{N}^{z})}
\mid 0 \rangle
\label{34tris}
\la

The corresponding Ricatti equation which fixes $S$ becomes in this context:

\ho
\lambda \sum_{{\bf j},\mu} \sigma_{{\bf j}}^{z} \sigma_{{\bf j},\mu}^{z}
+ E_{0} = -\sum_{ j=1}^{N} e^{\Delta_{ j}(S)}
\label{35}
\la

\noindent where

\ho
\Delta_{ j}(S)=  \frac{1}{2} [S(\sigma_{1}^{z},\cdots,\sigma_{j}^{z},
\cdots,\sigma_{N}^{z})
 - S(\sigma_{1}^{z},\cdots,-\sigma_{j}^{z},\cdots,\sigma_{N}^{z})
]
\label{35bis}
\la

The solution of this equation is rather cumbersome, even though in
principle it can be
obtained successively, step by step, as follows.
To start with, suppose we choose for the functional $S$ in
$(\ref{34tris})$ the expression:

\ho
S^{(1)}= - \alpha_{1} \sum_{{\bf j},\mu} \sigma_{{\bf j}}^{z}
\sigma_{{\bf j} + \mu}^{z}
\label{36}
\la

To this choice, it corresponds a state $\psi ^{(1)}= e^{-\frac{1}{2}
S^{(1)}} \mid 0\rangle$
which can be thought of as the ground state of a $D$-dimensional
FP-Hamiltonian:

\[H_{Ising,D}^{FP,(1)}\equiv H_{0} + U_{Ising,D}^{FP,(1)} -
E_{Ising,D}^{FP,(1)}\]
\ho
=- \sum_{j=1}^{N}(\sigma_{j}^{x} -
e^{\Delta_{j}(S^{(1)})})=\frac{1}{2}\sum_{j=1}^{N} {R_{j}^{(1)}}^{\dagger}
e^{\Delta_{j}(S^{(1)})} R_{j}^{(1)}
\label{36bis}
\la

\noindent where

\ho
R_{j}^{(1)}=  (e^{-\Delta_{j}(S^{(1)})}
\sigma_{j}^{x} - 1)
\label{36tris}
\la

\noindent and whose interaction potential $U_{Ising,D}^{FP,(1)}$ and energy
$E_{Ising,D}^{FP,(1)}$ are given by:

\[U_{Ising,D}^{FP,(1)} - E_{Ising,D}^{FP,(1)}= \sum_{j=1}^{N}
e^{\Delta_{j}(S^{(1)})}= \sum_{{\bf j}}
e^{-\alpha_{1}\sum_{\mu} \sigma_{{\bf j}}^{z}(\sigma_{{\bf j}+\mu}^{z}
+ \sigma_{{\bf j}-\mu}^{z})   }\]
\ho
=(\cosh \alpha_{1})^{2D} \sum_{{\bf j}} \prod_{\mu} (1-t_{1}\sigma_{{\bf
j}}^{z}\sigma_{{\bf j}
+\mu}^{z})
 (1-t_{1}\sigma_{{\bf j}}^{z}\sigma_{{\bf j}-\mu}^{z})
\label{37}
\la

\noindent with $t_{1}\equiv \tanh \alpha_{1}$.

Eq.(\ref{37}) yields for the energy $E_{Ising,D}^{FP,(1)}$:

\ho
E_{Ising,D}^{FP,(1)}=-N (\cosh \alpha_{1})^{2D}
\label{38}
\la

\noindent and for the $U_{Ising,D=1,2}^{FP,(1)}$ interaction Hamiltonians:

\ho
U_{Ising,D=1}^{FP,(1)}=-({\cosh \alpha_{1}})^{2} \sum_{j=1}^{N}(2 t_{1}
\sigma_{j}^{z}\sigma_{j+1}^{z} -
t_{1}^{2}\sigma_{j}^{z}\sigma_{j+2}^{z})
\label{39}
\la

\[U_{Ising,D=2}^{FP,(1)}=-({\cosh \alpha_{1}})^{4}
\sum_{{\bf j}}[2t_{1} \sum_{\mu}\sigma_{{\bf j}}^{z}\sigma_{{\bf j}+\mu}^{z}-
t_{1}^{2}\sum_{\mu}\sigma_{{\bf j}}^{z}\sigma_{{\bf j}+2\mu}^{z}\]
\[-2t_{1}^{2} \sigma_{{\bf j}}^{z}(\sigma_{{\bf j}+\mu_{1}+
 \mu_{2}}^{z} +  \sigma_{{\bf j}+\mu_{1}- \mu_{2}}^{z}  )
+t_{1}^{3} \sigma_{{\bf j}}^{z}(\sigma_{{\bf j}+\mu_{1}}^{z}
 \sigma_{{\bf j}-\mu_{1}}^{z}\sigma_{{\bf j}+\mu_{2}}^{z}\]
\ho
+\sigma^{z}_{{\bf j}+\mu_{1}} \sigma^{z}_{{\bf j}-\mu_{1}}
\sigma^{z}_{{\bf j}-\mu_{2}}
+ \mu_{1}\leftrightarrow \mu_{2})
+t_{1}^{4}\sigma^{z}_{{\bf j}+\mu_{1}} \sigma^{z}_{{\bf j}-\mu_{1}}
\sigma^{z}_{{\bf j}
+\mu_{2}}\sigma^{z}_{{\bf j}-\mu_{2}}]
\label{39bis}
\la

\vspace{0.5cm}
A glance at eqs.(\ref{39}) and (\ref{39bis}) shows that both
$U^{FP,(1)}_{Ising,D=1,2}$
contain a term
of the form $\sum_{{\bf j},\mu} \sigma_{{\bf j}}^{z} \sigma_{{\bf
j}+\mu}^{z}$.
In fact, this
is a general property of $U^{FP,(1)}_{Ising,D}$ in any dimension $D$.
Then, if we make the identification:

\ho
\lambda = 2(\cosh \alpha_{1})^{2D} \tanh \alpha_{1}
\label{40}
\la

\noindent we may say that
$U_{Ising,D}$ and $U^{FP,(1)}_{Ising,D}$ coincide to order $\lambda$ but
differ in order $\lambda^{2}$ and higher. On the other hand, if one
eliminates $\alpha_{1}$ in terms of $\lambda$ with the help of
(\ref{40}) and substitutes  back the result in (\ref{38}), one
immediately finds the energy
of the state $\psi^{(1)}$. In the $1D$ case, this energy reads:

\ho
E_{Ising,D=1}^{FP,(1)}= -\frac{N}{2}(1 + \sqrt{1 + \lambda^{2}})
\label{41}
\la

This expresion agrees to second order in $\lambda$ with the exact result
 which is given in the termodinamic limit
$N\longrightarrow \infty$  by \cite{Pfeuty1}:

\ho
e_{Ising,D=1}=-\lim_{N\rightarrow \infty} \frac{E_{Ising,D=1}}{N}=(1+\lambda)
\,\,F(-\frac{1}{2},\frac{1}{2};1; \frac{4\lambda}{(1+\lambda)^2})
\label{42}
\la

\noindent where $F(a,b;c;x)$ is the usual hypergeometric function \cite{Erdel}.

{}From eqs.(\ref{39}) and (\ref{39bis}) it is clear how to improve the first
order aproximation. One has to add to the exponent of the
ansatz those terms necessary to cancell the unwanted ones. In the $1D$
case that corresponds to the choice:

\ho
\psi^{(2)}= e^{\frac{1}{2}\sum_{j=1}^{N}[\alpha_{1}\sigma_{j}^{z}
\sigma_{j+1}^{z}
+\alpha_{2}\sigma_{j}^{z}\sigma_{j+2}^{z}]}
\label{43}
\la

The parameter $\alpha_{2}$ has been introduced as to eliminate the term
$\sum_{j=1}^{N}\sigma_{j}^{z}\sigma_{j+2}^{z}$ appearing in (\ref{39}).
Following the same steps as in (\ref{37}) we obtain for
the potential $U_{Ising,D=1}^{FP,(2)}$ and energy
$E_{Ising,D=1}^{FP,(2)}$ the equation:

\[U_{Ising,D=1}^{FP,(2)} - E_{Ising,D=1}^{FP,(2)}= (\cosh \alpha_{1}
\cosh \alpha_{2})^{2}\]
\ho
\times \sum_{j=1}^{N}(1 - t_{1}\sigma_{j}^{z}\sigma_{j+1}^{z})
(1 - t_{1}\sigma_{j}^{z}\sigma_{j-1}^{z})
(1 - t_{2}\sigma_{j}^{z}\sigma_{j+2}^{z})(1 - t_{2}\sigma_{j}^{z}
\sigma_{j-2}^{z})
\label{44}
\la

\noindent where

\ho
t_{a}\equiv \tanh \alpha_{a},\,\,\,\,\,\,\,\,a=1,2
\label{444}
\la

Now, by demanding $U_{Ising}$ agrees to order $\lambda^{2}$ with
$U_{Ising,D=1}^{FP,(2)}$ implies:

\[E_{Ising,D=1}^{FP,(2)}= -N (\cosh \alpha_{1} \cosh \alpha_{2})^{2}\]
\ho
\lambda = 2(\cosh \alpha_{1} \cosh \alpha_{2})^{2} t_{1}(1- t_{2})
\label{45}
\la
\[0= 2t_{2} - t_{1}^{2}\]

These equations can be easily solved for $E_{Ising,D=1}^{FP,(2)}$
as a function of $\lambda$, namely:

\ho
E_{Ising,D=1}^{FP,(2)}= -N \frac{\lambda}{t_{1}(\lambda)(2-t_{1}^{2}(\lambda))}
\label{45bis}
\la

\noindent where $t_{1}(\lambda)$  is the real and positive solution of
the polynomial

\ho
t_{1}^{4} + t_{1}^{2} + \frac{4t_{1}}{\lambda}-2=0\,.
\label{45tris}
\la

In fig.2 a plot of the values for minus the energy density
$e_{Ising,D=1}^{FP,(1)}$,
$e_{Ising,D=1}^{FP,(2)}$ and
$e_{Ising,D=1}$ is presented. Obviously the results improve
significantly as we increase the order of the aproximation.

This procedure can be repeated order by order by adding successively
terms of more complicated
structure in the exponential ansatz. The hope is that after a finite
number of steps the
process ends up. This is indeed  the case, as can be seen  for a $1D$ chain
with $N=4$ (see appendix B). We can say that the ground state solution
of the Hamiltonian (\ref{33})
can be constructed in a process which is similar to the one used in
perturbation
theory, but with two basic differences; firstly, it incorporates at
each step non perturbative
contributions to the energy and to the ground state; and secondly, the process
can be completed after a finite number of steps for finite chains.

\setcounter{equation}{0}
\section{A Fokker-Planck extension of the Ising model}

We have just described an attempt to reconstruct the exact ground state
of a 1-dimensional Ising model based in the choice of an exponential
ansatz. This is certainly feasible for small chains, as shown in
appendix B, but it becomes an awful task for chains already modest
in size. On the other hand, it is not our purpose to obtain in a
complicated manner what can be easily done by other means
\cite{Lieb}, \cite{Pfeuty1}. The Ising
model in $1D$ is an integrable system and can be solved by performing a
Jordan-Wigner transformation to fermion operators followed by a Bogoliubov
transformation.
 In higher dimensions, however, the model is probably not integrable
and that explains the difficulty in finding
an exact result, which quite likely does not exist.
One of the main interests to seek exact solutions of lattice models
consists in the study of their critical properties. Thus, if the model
under consideration comes from a discretization of a model in the
continuum, one
would like to recover that theory as the critical point of the corresponding
lattice model.
 In any case,
the choice of a particular hamiltonian is not so important as
the universality class to which it belongs to. So in this section
we  propose a FP-hamiltonian which
enjoies the same universality properties
 as the 1-dimensional ITF model, while remaining close to this latter
outside the critical region, in the sense that the energy
and other physical quantities do not deviate significantly.

Let us start by considering the most general 1-dimensional
ITF hamiltonian including all possible couplings involving
only two $\sigma^{z}$ matrices, namely:

\ho
H_{Ising}(\lambda_{1},\cdots,\lambda_{N})= -\sum_{j=1}^{N} \sigma_{j}^{x} -
\frac{1}{2} \sum_{j=1}^{N} \sum_{r=1}^{N-1}
\lambda_{r} \sigma_{j}^{z} \sigma_{j+r}^{z}
\label{46}
\la

\noindent where, for consistency:

\ho
\lambda_{r}=\lambda_{N-r}\,\,\,\,\,,\,\,\,\,\,r=1,\cdots,N-1
\label{46bis}
\la

Notice that the standard 1-dimensional ITF model corresponds to the
choice: $\lambda_{1}=\lambda_{N-1}=\lambda$ and $\lambda_{r}=0$ for $r\neq 1$
or $N-1$.

We want to find a FP hamiltonian which coincides with (\ref{46})
except for operators involving more than two $\sigma^{z}$-matrices, i.e:

\ho
U_{Ising}^{FP}(\lambda_{1},\cdots,\lambda_{N})-
U_{Ising}(\lambda_{1},\cdots,\lambda_{N})=
\sum_{p\geq 3}\sum_{i_{1},\cdots,i_{p}}
C_{i_{1},\cdots,i_{p}} \sigma_{i_{1}}^{z}\cdots\sigma_{i_{p}}^{z}
\label{47}
\la

According to the FP-philosophy, the knowledge of
$H_{Ising}^{FP}$ is equivalent to the knowledge of its
ground state, which we shall suppose to be given by:

\ho
\mid \psi (\alpha_{1},\cdots,\alpha_{N})\rangle =
e^{-\frac{1}{2}S(\sigma_{1}^{z},\cdots,\sigma_{N}^{z}) }\mid 0 \rangle
=    e^{\frac{1}{4}\sum_{r=1}^{N-1} \sum_{j=1}^{N}
\alpha_{r} \sigma_{j}^{z} \sigma_{j+r}^{z}} \mid 0 \rangle
\label{48}
\la

\noindent where $\alpha_{r}=\alpha_{N-r}$ are assumed to be real and
$\mid 0 \rangle$
satisfies (\ref{34bis}).

The discrete Ricatti equation reads for the ansatz
(\ref{48}):

\[E^{FP}+\frac{1}{2} \sum_{j=1}^{N} \sum_{r=1}^{N-1}
\lambda_{r} \sigma_{j}^{z} \sigma_{j+r}^{z}\]
\ho
-\sum_{p\geq 3}\sum_{i_{1},\cdots,i_{p}}
C_{i_{1},\cdots,i_{p}} \sigma_{i_{1}}^{z}\cdots\sigma_{i_{p}}^{z}=
-\sum_{j=1}^{N}e^{\Delta_{j}(S)}
\label{49}
\la

\noindent where

\ho
\Delta_{j}(S)= -\sum_{r=1}^{N-1} \alpha_{r} \sigma_{j}^{z}
\sigma_{j+r}^{z}
\label{49bis}
\la

The $\sigma_{j}^{z}$ operators in eq.(\ref{49}) can in fact be
replaced by variables $\sigma_{j}$ taking just two values $\pm 1$.
Then if we multiply both sides of (\ref{49}) by a convenient
number of $\sigma_{j}$'s and  subsequently we sum over
the $2^{N}$ possible choices of $\sigma_{j}$'s we obtain
for the energy $E^{FP}$ and  the coupling constants $\lambda_{r}$

\ho
E^{FP}= -\frac{N}{2^{N}} \sum_{\{\sigma_{1},\cdots,\sigma_{N}\}}
e^{\Delta_{0}(S)}
\label{50}
\la

\ho
\lambda_{r}= -\frac{1}{2^{N}} \sum_{\{\sigma_{1},\cdots,\sigma_{N}\}}
(\sum_{j=1}^{N} \sigma_{j}\sigma_{j+r})
e^{\Delta_{0}(S)}
\label{50bis}
\la

Here the subindex $0$ in $\Delta_{0}(S)$ refers
to the site ``$0$'' (= site ``$N$'') of the chain. It is straightforward to
prove that

\ho
e^{\Delta_{0}(S)}= \prod_{r=1}^{N-1} \cosh \alpha_{r}
(1- t_{r} \sigma_{0}\sigma_{r})
\label{51}
\la

\noindent where $t_{r}\equiv \tanh \alpha_{r}$. If we now
introduce (\ref{51}) in (\ref{50})-(\ref{50bis}) and define:

\ho
1=- t_{0}
\label{53bis}
\la

\noindent we obtain:

\ho
e_{N}^{FP}\equiv \frac{-E^{FP}}{N}= \prod_{r=1}^{N-1} \cosh \alpha_{r}
\label{52}
\la

\ho
\frac{\lambda_{r}}{e_{N}^{FP}}=-
\sum_{n=0}^{N-1}
t_{n}t_{r-n}\,\,\,\,\,,\,\,\,\,\,(r=1,\cdots,N-1)
\label{53}
\la

\noindent where $e_{N}^{FP}$ is minus the FP-energy density of
a chain of length $N$.

The structure of eqs.(\ref{52})-(\ref{53}) has already
emerged in, for example, eq.(\ref{45}) (see also eqs.
(\ref{13b})-(\ref{14b}) of appendix B). These equations
relate the $t_{n}$ (or the $\alpha_{n}$) to the coupling
constants $\lambda_{r}$ and, therefore, one could in principle
relate the  FP-energy $E^{FP}$ with the $\lambda$'s. In this whole process,
the parameters $C_{i_{1},\cdots,i_{p}}$ which stablish
the difference between our model and the pure Ising model
(\ref{46}) play no role. In fact, they can be computed a posteriori
in terms of the $\alpha$'s. This means that our prescription
to associate a FP-model to the Ising is well defined.

The next step is the resolution of eqs.(\ref{52})-(\ref{53}). For that,
let us introduce the Fourier transformed ${\hat{t}}_{n}$ of the $t_{n}$:

\ho
t_{n}=\frac{1}{\sqrt{N}} \sum_{m=0}^{N-1} e^{\frac{2 \pi i n m}{N}}
{\hat{t}}_{m}
\label{54}
\la

\noindent and consequently eq.({\ref{53}) becomes:

\ho
\frac{\lambda_{r}}{e_{N}^{FP}}=-\sum_{m=0}^{N-1} e^{\frac{2 \pi i rm}{N}}
({\hat{t}}_{m})^{2}
\label{55}
\la

Then, if we apply an inverse Fourier transformation to (\ref{55}) we get the
${\hat{t}}_{n}$'s in terms of the $\lambda$'s and a parameter $C_{N}$, namely:

\ho
{\hat{t}}_{m}=-\frac{1}{\sqrt{N}}\varepsilon_{m}( C_{N} -
\sum_{r=1}^{N-1} e^{\frac{-2 \pi i r m}{N}}
\,\,\frac{\lambda_{r}}{e_{N}^{FP}})^{1/2}
\label{56}
\la

\noindent where $\varepsilon_{m} = \varepsilon_{N-m} = \pm 1$  and

\ho
C_{N}= \sum_{m=0}^{N-1} ({\hat{t}}_{m})^{2} =  \sum_{n=0}^{N-1} t_{n}^{2}
\label{56bis}
\la

\noindent is a real constant which  can be fixed by imposing the eq.
(\ref{53bis}):

\ho
1=-\frac{1}{\sqrt{N}} \sum_{m=0}^{N-1}  {\hat{t}}_{m}=
\frac{1}{N}\sum_{m=0}^{N-1}\varepsilon_{m}( C_{N} -  \sum_{r=1}^{N-1}
e^{\frac{-2 \pi i r m}{N}}
\,\,\frac{\lambda_{r}}{e_{N}^{FP}})^{1/2}
\label{57}
\la

One can easily realize that the trivial solution for the case $\lambda_{r}=0,
\,\,\forall r$, corresponds to:

\[t_{n}= -\delta_{n,0}\,\,\,\,\,\,,\,\,\,\,\,\,{\hat{t}}_{m}=
-\frac{1}{\sqrt{N}}\]
\ho
C_{N}=1\,\,\,\,\,\,,\,\,\,\,\,\,\varepsilon_{m} = 1
\label{58}
\la

The presence of interaction modifies these values; for  convenience, however,
we shall continue maintaining $\varepsilon_{m} = 1$.

When $\lambda = \lambda_{1}= \lambda_{N-1}$, ($\lambda_{r}=0$, otherwise),
 one recovers the usual ITF model (\ref{33}).
Eqs.(\ref{54}) and (\ref{57})
becomes now:

\ho
t_{n}=-\frac{1}{N}\sum_{m=0}^{N-1}e^{\frac{2 \pi i n m}{N}} ( C_{N} -
\frac{2\lambda}{e_{N}^{FP}} \cos(\frac{2\pi m }{N}))^{1/2}
\label{59}
\la

\ho
1=\frac{1}{N}\sum_{m=0}^{N-1} ( C_{N} -
\frac{2\lambda}{e_{N}^{FP}} \cos(\frac{2\pi m }{N}))^{1/2}
\label{59bis}
\la

Notice that  (\ref{59bis}) is nothing but the $n=0$ case of (\ref{59}).

We shall look for solutions of (\ref{59})and (\ref{59bis}) for which
the $t_{n}$ are
real quantities. This is guaranted if we impose the  constraint:

\ho
0\leq u_{N} \equiv \frac{2 \lambda}{ C_{N} e_{N}^{FP}} \leq 1
\label{59tris}
\la

The solution of (\ref{59})and (\ref{59bis}) involves, in principle, four steps:
i) to express $C_{N}$ as a function of $\lambda$ and $e_{N}^{FP}$,
ii) to rewrite $t_{n}$ in terms of $\lambda$ and $e_{N}^{FP}$,
iii) to introduce the $t_{n}$ in eq.(\ref{52}) as to obtain an equation for
$e_{N}^{FP}$ as a function of  $\lambda$, and  iv) to solve this equation.

Fortunately, this rather complicated elimination process
 can be explicitely accomplished in two extreme
cases according as: a) $u_{N}=1$ and $N$ arbitrary or
  b) $u_{N}<1$ but $N \longrightarrow \infty$. Next we shall describe
our results
leaving the details and derivations to the Appendix C.

\vspace{0.5cm}

\underline {{\bf a) Case $u_{N}=1$}}

The values of $t_{n}$ and $e_{N}^{FP}$ are given by:

\ho
t_{n}=\frac{(\sin \frac{\pi}{2N})^{2}}{\sin\frac{\pi}{N}(n +\frac{1}{2})
\sin\frac{\pi}{N}(n -\frac{1}{2})}
\label{61bis}
\la


\ho
e_{N}^{FP} = \frac{\sqrt{2}}{N \sin \frac{\pi}{2N}
\sin \frac{N \theta_{N}}{2}}
\label{63}
\la

\noindent where the angle $\theta_{N}$ in (\ref{63}) is defined as:

\ho
\cos \theta_{N}= 2 \cos(\frac{\pi}{N}) -1
\label{62penta}
\la

If we take the limit
$N\longrightarrow \infty$ in eqs.(\ref{61bis}) and (\ref{63}) we deduce
that the model exhibits critical behaviour. First of all, from
eq.(\ref{61bis}) we obtain

\ho
\lim_{N\rightarrow \infty} t_{n} = \frac{1}{4 n^{2} -1}
\label{62m}
\la

\noindent and hence for $n$ large the ``amplitudes'' $t_{n}$ have a
scaling behaviour. It is curious to observe that the value of $t_{n}$
given by eq.(\ref{62m})
coincides with the correlator of two $\sigma^{x}$'s matrices in the
standard ITF model at the critical point \cite{Pfeuty1}:

\ho
\langle \sigma_{j}^{x} \sigma_{j + n}^{x} \rangle -
\langle \sigma_{j}^{x}  {\rangle}^{2} = \left( \frac{2}{\pi} \right)^{2}
\frac{1}{4 n^{2} -1}
\label{62mmm}
\la

We may wonder whether these two results are related.

\vspace{0.5cm}
Another indication of criticality arises from the $1/N^{2}$ correction to
the ground state energy density, which is given by:

\ho
e_{N}^{FP} =
 e_{\infty}^{FP} [1 + \frac{{\pi}^{2}}{24 N^{2}}
(1 - \frac{\pi}{\sqrt{2}} \cot (\frac{\pi}{\sqrt{2}}))+ O(\frac{1}{N^{3}})]
\label{63m}
\la

\noindent where

\ho
e_{\infty}^{FP}= \frac{2 \sqrt{2}}{\pi \sin (\pi/\sqrt{2})}\cong
\frac{3.554}{\pi}
\label{63mm}
\la

Recall that $e_{N}^{FP}$ is minus the energy density of the ground state.

It is well known that in a $1D$ conformal invariant theory this correction
satisfies the Cardy formula:

\ho
e = e_{\infty} + \frac{\pi c v_{s}}{6 N^{2}}
\label{64m}
\la

\noindent where $v_{s}$ is defined through the dispersion relation of
the elementary
excitation above the ground state, i.e. $\omega(k) \approx v_{s} k$,
and $c$ is the central extension of the Virasoro algebra underlying the
corresponding conformal field theory.

Compairing eqs. (\ref{63m}) and (\ref{64m}) we learn

\ho
c\,v_{s} = \frac{\pi}{4} e_{\infty}^{FP}(1 - \frac{\pi}{\sqrt{2}}
\cot (\frac{\pi}{\sqrt{2}}))
\label{61tris}
\la

Hence we would need to know the value of $v_{s}$ in order to derive
the central extension $c$
(see section 5 for some considerations concerning this point). Our expectations
are that $c= \frac{1}{2}$, in which case we would be actually describing
an Ising model;
but we shall arrive to this result by studying other quantities, such
as the critical exponents.

It is amusing to observe that the value of the critical energy
density (\ref{63mm}) is comparable with the one of the ITF model:

\ho
e_{\infty}^{ITF}=\frac{4}{\pi}
\label{65}
\la

To complete this comparison, let us recall that in 1D the critical
value of $\lambda$ is
given by $\lambda_{c}^{ITF}=1$, whereas in our FP-ITF model we find

\ho
\lambda_{ c}^{FP}=\frac{\pi}{4 \sqrt{2} \sin (\pi / \sqrt{2})} \cong 0.6979
\label{64}
\la

\vspace{0.5cm}

\underline{{\bf b) Case $u_{\infty}<1$}}

{}From the discussion of the previous paragraph we expect that this case
will account for the non critical regime of the model. The expressions for
the $t_{n}$ and the energy $e^{FP}_{\infty}$ are given here by (see Appendix C
for computation details ):

\ho
t_{n} = \frac{1}{\pi 2^{5/2}} \,\,\left[ \frac{2 \lambda_{R}}
{(1 + \lambda_{R})^{2}}\right] ^{n}\,\, \frac{\Gamma(\frac{n}{2} - \frac{1}{4})
\Gamma(\frac{n}{2} + \frac{1}{4})}{\Gamma(n + 1)}\,\,
\frac{F(n - \frac{1}{2}, n + \frac{1}{2}, 2n + 1; \frac{4 \lambda_{R}}
{(1 + \lambda_{R})^{2}})}{F(-\frac{1}{2}, \frac{1}{2}, 1; \frac{4 \lambda_{R}}
{(1 + \lambda_{R})^{2}})}
\label{65p}
\la

\ho
e_{\infty}^{FP} = \frac{\lambda}{\lambda_{R}}
\left[ (1 + \lambda_{R}) \,\,F (-\frac{1}{2}, \frac{1}{2}, 1; \frac{4
\lambda_{R}}{(1 + \lambda_{R})^{2})}\right]^{2}
\label{66p}
\la

\noindent where we have introduced a new variable $\lambda_{R}$ through:

\ho
u_{\infty} = \frac{2 \lambda_{R}}{1 + \lambda_{R}^{2}}
\label{67p}
\la

Equations (\ref{65p}) and (\ref{66p}) make clear that $\lambda_{R}$ rather
than $\lambda$ is the effective coupling constant which governs the behaviour
of the ground state. In a certain sense, $\lambda_{R}$ results from
the dressing of the ``bare''
coupling constant $\lambda$ by all the interactions present in the
FP-Hamiltonian.
The explicit form of $\lambda$ as a function of $\lambda_{R}$
is very complicated and it will not be needed; in any case it can be
obtained by
comparing eq.(\ref{67p}) with the eq.(\ref{c20}) of Appendix C for
$e^{FP}_{\infty}$
as a function of the $t_{n}$'s. In fact, it is not difficult to show that

\ho
\lambda = \lambda_{R} (1 - \frac{1}{2} \lambda_{R}^{2} + O(\lambda_{R}^{4}))
\label{68p}
\la

\noindent and

\ho
\lim_{\lambda_{R} \rightarrow 1}\left(\frac{d \lambda}{d \lambda_{R}}\right)=0
\label{69p}
\la

 Making use of (\ref{69p}) one sees that $\frac{\partial
e^{FP}_{\infty}}{ \partial \lambda_{R}}$
is not singular at the critical value $\lambda_{R}=1$. However, the
``specific heat''
defined as the second derivative $\frac{ {\partial}^{2} e^{FP}_{
\infty}}{ \partial
{\lambda_{R}}^{2}}$ does actually exhibit a logarithm divergence

\ho
\frac{ {\partial}^{2} e^{FP}_{\infty}}{ \partial
{\lambda_{R}}^{2}} \simeq -\frac{8 \lambda_{c}}{\pi ^{2}}
\log (1 - \lambda_{R})
\label{70p}
\la

This is precisely the same logarithm singular behaviour that one would
expect from the original
ITF model \cite{Pfeuty1}. This implies that the critical exponent
$\alpha$ defined through
$\frac{ {\partial}^{2} e}{ \partial
{\lambda}^{2}} \simeq (1 - \lambda)^{\alpha}$ is the same for the two
models, i.e. $\alpha = 0$.

Let us concentrate now on the behaviour of the $t_{n}$ for\,\, $n$ large. In
 the critical case
$\lambda_{R} = 1$, we have seen that they satisfy the scaling law
(\ref{62m}); if
$\lambda_{R} < 1$ we obtain instead

\ho
t_{n} \sim \frac{1}{n^{3/2}} \,\,\lambda_{R}^{n}
\,\,\,\,\,\,\,\,{\rm for}\,\,\, n >> 1,\,\,\,\,\lambda_{R}<1
\label{71p}
\la

 \noindent and so eq.(\ref{71p}) allows one to define a correlation
length $\xi(\lambda_{R})$
through $t_{n} \sim \frac{1}{n^{3/2}}\,e^{-\frac{n}{\xi}}$ as

\ho
{\xi}^{-1} ( \lambda_{R}) = - \log \lambda _{R} \simeq (1 - \lambda_{R}),
\,\,\,\,\,\lambda_{R} \simeq 1
\label{72p}
\la

\noindent which implies in particular that the critical exponent $\nu$ is
equal to 1, just like in the standard ITF model. This is again another
 indication that the model we are
considering belongs to the same universality class as the standard ITF model.

\setcounter{equation}{0}
\section{A change of variables: excited states}

Until now we have considered the construction of the ground state of a lattice
Hamiltonian in the spirit of the FP procedure; at this point, it seems natural
to wonder what can tell us the method about the rest of the eigenstates of
the Hamiltonian. This is,
of course, important both for the study of gaps in the spectrum
as for the dispersion relation of elementary excitations.

In order to fix ideas, let us start by posing the problem in the context of
Quantum Mechanics. Hitherto, we have been using the fact that a wave
function of the form
$\psi_{0} (x) = e^{- S(x)}$,
 \,\,$x\in {\bf R}$, is a ground state of the FP-Hamiltonian:

\ho
H^{FP}=\frac{1}{2}\left[- \frac{d^2}{d x^2} + (\frac{d S}{d x})^2 -
\frac{d^2 S}{d x^2}\right]
\label{e1}
\la

If, for example, we take $S(x)=\frac{1}{2} x^{2}$, then  $H^{FP}$
coincides with
 the number operator for the harmonic oscillator
$(H^{FP}=  a^{\dagger} a ={\cal N},\,\, [a, a^{\dagger}]=1)$.
If we want to study the excited states of (\ref{e1}), we could make the ansatz:
$ \psi (x) = A(x) e^{- S(x)}$, which can be thought of as a change of
variables from
$\psi (x)$ to $A (x)$ since the function $S(x)$ is known a priori.
The eigenvalue problem for $\psi (x)$ becomes then:

\ho
H^{FP} \psi (x)= \Delta E \psi (x)\,\,\,\,\,\,\Longrightarrow
 \,\,\,\,\,\,H^{RFP} A (x)= \Delta E\,\, A (x)
\label{e2}
\la

\noindent where

\ho
H^{RFP}=\frac{1}{2}\left[- \frac{d^2}{d x^2} + 2 \frac{d S}{d x}\,\,
\frac{d}{dx} \right]
\label{e3}
\la

\noindent and $\Delta E$ represents the difference of energies
between the excited state
$\psi(x)$ and the ground state $\psi_{0}(x)$ -whose energy is actually zero.
 We shall call $H_{RFP}$ the reduced Fokker-Planck
Hamiltonian. The ground state solution using the new variable $A(x)$
corresponds to $A(x)=Cte$. In particular, for the harmonic oscillator,
eq.(\ref{e2}) reads:

\ho
-\frac{1}{2} \frac{d^2 A}{d x^2} + x \frac{d A}{d x} = \Delta E\,\, A
\label{eee3}
\la

\noindent which becomes a confluent hypergeometric equation after
the substitution
$y=x^{2}$, and whose normalized solutions
are  given by the well-known Hermite polynomials. It is worth
noticing that $H^{RFP}$ is not
an hermitean operator with respect to the measure $\mu= d x$, but it is so
for the measure $\mu= e^{-2 S(x)} d x$.

Next, we shall try to generalize the previous methods to more
complicated Hamiltonians.
More precisely, the problem we face is the construction of the
 spectrum of the FP-Hamiltonian associated to (\ref{46}), namely:

\ho
 H^{FP}= - \sum_{j=1}^{N} [\sigma_{j}^{x} - e^{\Delta_{j} (S)}]
=- \sum_{j=1}^{N} [\sigma_{j}^{x} - e_{N}^{FP} \prod_{r=1}^{N-1}
(1 - t_{r} \sigma_{j}^{z} \sigma_{j+r}^{z})]
\label{e4}
\la

\noindent whose ground state $\psi_{0}$ was defined in (\ref{48}):

\ho
\mid \psi_{0}\rangle =
  \exp(\frac{1}{4}\sum_{r=1}^{N-1} \sum_{j=1}^{N}
\alpha_{r} \sigma_{j}^{z} \sigma_{j+r}^{z}) \mid 0 \rangle
\label{e5}
\la

In (\ref{e4}) $e_{N}^{FP}$ represents minus the FP-energy density given
by (\ref{52}) and $t_{r}\equiv \tanh \alpha_{r}$. In principle, this
ground state may not be unique. For example, this would happen if
the Hamiltonian (\ref{e4}) enters into an ordered region characterized
by ground states with opposite non-vanishing magnetization. We shall
not discuss this possibility here.

The change of variables adequate for our purposes
corresponds to:

\ho
\mid \psi \rangle = A(\sigma_{1}^{z},\cdots,\sigma_{N}^{z}) \mid
\psi_{0} \rangle
\label{e6}
\la

\noindent where $A(\sigma_{1}^{z},\cdots,\sigma_{N}^{z}) $
is the most general operator involving $\sigma^{z}$'s. With the notations of
appendix B:

\ho
A(\sigma_{1}^{z},\cdots,\sigma_{N}^{z}) = \sum_{a=1}^{2^{N}} A_{I_{a}}
\sigma_{I_{a}}^{z}
\label{e7}
\la

\noindent where in contrast to the appendix, we shall allow in
$I_{a}$ the presence of
any number of $1$'s, independently of their parity. The eigenvalue
equation (\ref{e2})
reads now:

\ho
H^{RFP} \mid A \rangle = \Delta E \mid A \rangle
\label{e8}
\la

\noindent where

\ho
\mid A \rangle  =A(\sigma_{1}^{z},\cdots,\sigma_{N}^{z}) \mid 0 \rangle \equiv
\sum_{a=1}^{2^{N}} A_{I_{a}} \mid I_{a} \rangle
\label{e9}
\la

\noindent and

\ho
H^{RFP}= \sum_{j=1}^{N} e^{\Delta_{j}(s)} ({\bf 1} - \sigma_{j}^{x})=
e^{FP}_{N} \sum_{j=1}^{N}\prod_{r=1}^{N-1}({\bf 1}- t_{r} \sigma_{j}^{z}
\sigma_{j+r}^{z})
({\bf 1}- \sigma_{j}^{x})
\label{e10}
\la

For reasons will be evident latter, it will be of some convenience for
us to introduce the parameter:

\ho
\varepsilon \equiv \frac{\Delta E }{2 e^{FP}_{N}}
\label{e14}
\la

\noindent and in what follows we will also assume that  $H^{RFP}$ is
divided by $2 e^{FP}_{N}$ to give ${\hat H}^{RFP}$. Notice that in
the absence of interaction
 (i.e $t_{r}=0, \,\, \forall r$), we have:

\ho
{\hat H}^{ (0)\,\,RFP}= \sum_{j=1}^{n} \frac{{\bf 1}-\sigma_{j}^{x}}{2}
\equiv {\cal N}
\label{e11}
\la

\noindent and so ${\hat H}^{0\,\,RFP}$ coincides with the number
operator $\cal N$,
i.e, the operator which in this context counts the number $n_{I_{a}}$
of $1$'s of a given state $\mid I_{a} \rangle$:

\ho
{\cal N} \mid I_{a} \rangle= n_{I_{a}} \mid I_{a} \rangle
\label{e12}
\la

An interesting property of the operator ${\hat H}^{RFP}$ is

\ho
\sum_{a=1}^{2^{N}} \langle I_{a} \mid {\hat H}^{RFP} \mid I_{b}
\rangle = \prod _{r=1}^{N-1}
(1 - t_{r}) \,\,n_{I_{b}}
\label{e13}
\la

This equation suggests some kind of connection between the evolution
described by the Hamiltonian ${\hat H}^{RFP}$ and a Markov process.
As a consequence of (\ref{e13}) we find

\ho
\varepsilon = \prod_{r=1}^{N-1}(1 - t_{r})
\,\,\,\frac{\sum_{a=1}^{2^{N}} n_{I_{a}} A_{I_{a}}}
{\sum_{a=1}^{2^{N}} A_{I_{a}}}
 \label{e15}
\la

A useful way of reexpressing ${\hat H}^{RFP}$ is

\[
{\hat H}^{RFP}(\{t_{r}\}) = -\sum_{r=0}^{N-1} t_{r} \Lambda_{r} +
\sum_{1\leq r_{1} < r_{2}\leq N-1 }t_{r_{1}} t_{r_{2}} \Lambda_{r_{1},r_{2}}+
\cdots \]
\[\cdots + (-1)^{l} \sum_{1\leq r_{1} < r_{2}<\cdots r_{l}\leq N-1 }
t_{r_{1}}\cdots t_{r_{l}} \Lambda_{r_{1},\cdots,r_{l}}+\cdots\]
\ho
- t_{1}t_{2}\cdots t_{N-1}\,\, \Lambda_{1,2,\cdots,N-1}
\label{e16}
\la

\noindent where the operators $\Lambda$'s are defined as

\ho
\Lambda_{r_{1},\cdots,r_{l}}= \sum_{j=1}^{N} (\sigma_{j}^{z})^{l}\,\,
\sigma_{j + r_{1}}^{z}\cdots
\sigma_{j + r_{l}}^{z}\,\,\frac{1- \sigma_{j}^{x}}{2}\,\,\,\,\,\,,\,\,\,\,\,\,
\Lambda_{0}\equiv \cal N
\label{e17}
\la

Notice that the first sum in (\ref{e16}) starts at $r=0$ and so reproduce
 the operator
$\cal N$ (recall that $t_{0}=-1$).

The action of $\Lambda$ on a generic state $\mid I_{a} \rangle =
\mid \mu_{1},\cdots,\mu_{N}\rangle$, ($\mu_{i} = 0\,\, {\rm or}\,\, 1,
{\rm mod}(2)$) is given by (see fig.3):

\ho
\Lambda_{r_{1},\cdots,r_{l}} \mid \mu_{1},\cdots,\mu_{N}\rangle =
\sum_{j=1}^{N} \delta _{1, \mu_{j}} \mid \cdots, \mu_{j} + l,
\cdots,\mu_{j+r_{1}} + 1, \cdots,\mu_{j+r_{l}} + 1,\cdots \rangle
\label{e18}
\la

The diagonalization of the Hamiltonian (\ref{e16}) seems, a priori, not much
easier than the diagonalization of the Hamiltonian (\ref{e4}).
However, we have developed a real renormalization group approach
which enables us to
deal with this apparent difficulty; but the results concerning this
point will be presented
elsewhere \cite{Fernan}. For the moment, and with the
purpose of getting some insight into the Hamiltonian (\ref{e16}),
let us consider only
the first term in it which is linear in the $t$'s, namely:

\ho
{\hat H}^{ (1)\,\,RFP}= - \sum_{r=0}^{N-1} t_{r} \Lambda_{r}
\label{e19}
\la

It is easy to see then that the state:

\ho
\mid \psi_{m} \rangle = \sum _{j=1}^{N} e^{2 \pi i m j/N} \,\,\mid j \rangle
\label {e20}
\la

\noindent where $\mid j \rangle=\mid 0\cdots 0\stackrel{j}{1} 0\cdots 0
\rangle$ is an eigenstate of ${\hat H}^{(1)\,\,RFP}$:

\ho
{\hat H}^{(1)\,\,RFP} \mid \psi_{m} \rangle = - \sqrt{N} \,\,
{\hat{t}}_{m}\mid \psi_{m} \rangle
\label{e21}
\la

Hence, eq.(\ref{56}) can be used to obtain
the dispersion relation for a plane wave of momentum $k=\frac{2 \pi m}{N}$

\ho
\Delta E^{FP}(k)= 2e^{FP}_{N} \left( C - \frac{2 \lambda \cos k}{e^{FP}_{N}}
\right)^{1/2}
\label{e22}
\la

The gap is obtained for $k=0$ and vanishes in the critical point, i.e when
$u_{N}=1$. Moreover, for $k$ small
the dispersion relation (\ref{e22}) is linear in $k$:

\ho
\Delta E^{FP}(k)\simeq \frac{\pi}{2}\,\,e^{FP}_{\infty} k
\label{e23}
\la

\noindent whence we deduce a speed of sound $v_{s}^{(1)}=
\frac{\pi}{2}\,e^{FP}_{\infty}$.
This confirms our expectations that the value $u_{N}=1$ actually corresponds
to a critical point.
Unfortunately, the speed of sound we derive from eq.(\ref{e23}) is not
the speed of sound
of the excitations of the whole Hamiltonian (\ref{e16}).
 In fact if we take for granted that $c=1/2$ is the value of
the central extension of this model, then the speed of sound for the
Hamiltonian (\ref{e19}) should be given by:

\ho
v_{s}^{(1)} = \frac{\pi}{2} e^{FP}_{\infty}(1 - \frac{\pi }{\sqrt{2}}
\cot (\frac{\pi }{\sqrt{2}}))
\label{e25}
\la

This equation suggests that the role of the additional terms in the
Hamiltonian (\ref{e16}) is simply to redefine the value of $v_{s}^{(1)}$.

In section 4 we introduced an effective coupling constant $\lambda_{R}$
in terms of which the energy $e_{\infty}^{FP}$ displayed a critical behaviour.
 If we now rewrite eq.(\ref{e22}) as a function of $\lambda_{R}$ we obtain

\ho
\Delta E^{FP}(k)= 2 \sqrt{\frac{\lambda e_{\infty}^{FP}}{\lambda_{R}} }
\,\,(1 + {\lambda}_{R}^{2} - 2 \lambda_{R} \cos k)^{1/2}
\label{e26}
\la

It is noteworthy to compare this relation with the exact
dispersion relation of the standard ITF model. As it is shown in \cite{Pfeuty1}

\ho
\Delta E_{ITF}^{exact}(k)= 2 (1 + {\lambda}^{2} - 2 \lambda \cos k)^{1/2}
\label{e27}
\la

The evident similarity between eqs.(\ref{e26}) and (\ref{e27}) is again another
confirmation that the model we are describing is very close to the
usual Ising model; indeed,
we believe they are identical at the critical point.

\setcounter{equation}{0}
\section{Conclusions and final remarks }

This paper represents a first step in the application of the Fokker-Planck
techniques to the study of quantum lattice Hamiltonians. The main idea is to
 approximate the Hamiltonian of interest $H$ by a FP-Hamiltonian $H^{FP}$
which shares with $H$ a certain number of properties. Firstly, we
have considered
FP-Hamiltonians which agree with $H$ to first and second order in
perturbation theory.
Next, and more interesting, we have built up a FP-Hamiltonian
 belonging to the same
universality class as $H$. In both cases the advantage of working with
a FP-Hamiltonian is that
we know the exact ground state of this latter. In the example we have
considered, i.e. the ITF
model, we have been able to solve the equations fixing the
ground state parameters in terms of the coupling constant,
and that makes feasible
the study of the critical behaviour of the model.
\vspace{3mm}

 Below we list some issues we believe deserve further investigation.
\vspace{8mm}

\underline{{\bf Modelistic:}}

\begin {itemize}

\item {\em The Ising model for $D >1$}.  Sections 4, 5 and 6 can automatically
be extended  to ITF models in higher dimensions. As a matter of fact,
increasing the space dimension
does not represent a priory a barrier to the FP-techniques, as is shown
in appendix D.
Our feeling is, however, that realistic results would require of
 exponential ansatz involving more
than two $\sigma^{z}$ operators. In this sense, the analysis of the
$D=2$ ITF model
still remains as a challenge in view of its connection
with the $3D$ classical Ising model.

\item {\em ${\bf Z_{N}}$-gauge models in $D=2$ and 3}. Here we can
inmediately apply the
 FP-procedure developed in the paper to construct the ground state in
the confining phase;
 in $2D$ this phase is dual to the ordered phase of the $2D$ ITF
model \cite{Rabi}.

\item{\em The  Potts model}.  FP general machinery of sections 2 can be
easily applied
to the case of the
1D Potts model \cite{Fernan}. The Potts model is a fruitful generalization
of the Ising model and has been received much attention in the context
of statistical mechanics. In the notation of section 2,
the k-state
$1D$ Potts Hamiltonian is defined by:

\[H_{Potts}(\lambda)= H_{0} + \lambda U_{Potts} (Z_{1},\cdots,Z_{N})\]
\ho
= -\sum_{n=1}^{k-1}\sum_{j=1}^{N}  X_{j}^{n} - \lambda
 \sum_{n=1}^{k-1} \sum_{j=1}^{N}(Z_{j} Z_{j+1}^{\dagger})^{n}
\label{70aa}
\la

\noindent where the operators $X$ and $Z$ verify
$XZ=e^{\frac{2\pi i}{k}} ZX$ and
$XZ^{\dagger}=e^{-\frac{2\pi i}{k}}Z^{\dagger}X$. The Hamiltonian
(\ref{70aa}) constitutes an example of a traslational
${\bf Z}_{k}$ invariant and hermitic operator, which includes the
Ising Hamiltonian as a particular case ($k=2$).

\item{\em Fermion models and Heisenberg models}. In the example considered
in this paper, the operators $V_{I}$ entering in the exponential of the ansatz
commuted among themselves. This property facilitated a lot the
computations. However,
this is not the case for models with fermions or spins \cite{Delgado}.
Hence, the question of a convenient
implementation of  the FP-techiques so as to include fermions still
remains to be done.

\item{\em Degenerate ground states}. The general FP-construction
introduced in section 2
applies also when the unperturbed hamiltonian $H_{0}$ has degenerated
ground states, as is the case of the ITF model in the ordered phase;
here the computation
goes in parallel to the well-known degenerated perturbation theory of
Quantum Mechanics.

\underline{{\bf Universality:}}

\item Apart from the critical exponent $\alpha$ and $\nu$ computed in
this paper,
one would like to determine the rest of exponents, specially
the fractional ones. In fact, only after this determination had been
accomplished,
could a complete agreement with the critical
Ising model be claimed.

\item If our statement concerning the universality class of
the FP-Hamiltonians is correct, then we have a method to compute
the exact values of critical exponents. One may wonder
how the scenario would be like in dimensions higher than one, where
non trivial integrable models do not exist.

\item FP-procedure seems to throw a light on the way we could compute
the correlation functions of some operators. For that purpose,
we could use the ground state energy as a
generating function of correlations along the lines of
the Hellmann-Feynman theorem.

\item A fundamental question concerning the FP-aproach undertaken in this paper
is how to find a ``minimal FP-ansatz'' capable of reproducing the same
critical behaviour of the original
Hamiltonian $H$. We do not have a neat answer to this question yet; however
common sense recommends to begin with ``simple'' ansatzs and from these
to construct more complicated ones. The final objective must be, of course,
 to capture the essential physics of the model under consideration.

\underline{{\bf Integrability:}}

\item An interesting open question is whether the FP-Hamiltonians
are integrable or not, since this would explain some of the magic
these Hamiltonians enjoy. If so, one would like to know whether or not
there exists a kind of Yang-Baxter equation underlying or responsable for such
a behaviour.
\end{itemize}

\vspace{20 mm}

\noindent {\Large{\bf Acknowledgments}}
\vspace{5pt}

We would like to thank E. Brezin, J.G. Carmona, J.G. Esteve,
 C. G{\'o}mez, M.Halpern,
M.A. Martin-Delgado and M.A.H. Vozmediano for useful discussions. This work
has been  partially supported by the CICYT grant PB92-1092 (G.S.).

\newpage

\newcommand{\Appendix}[1]{\appendix{#1}\setcounter{equation}{0}}

\appendix

\Section{ A $q$-deformation of the FP-quantum
 mechanical Hamiltonians}

In Section 2 we were concerned with a Fokker-Planck
approach to Hamiltonians which appear in continuous or lattice Quantum
Field Theories. We devote this appendix to show that
most of the formulae appearing there  have an analogue  in
ordinary  Quantum Mechanics. For pedagogical reasons, let us concentrate on
$1D$.

To begin with, let us write down the analogous to eqs. (\ref{3}),
 (\ref{4}) and (\ref{5}) for a wave function $\psi (x) = e^{- S(x)}$,
 \,\,$x\in {\bf R}$:

\[H^{FP}=\frac{1}{2}[- \frac{d^2}{d x^2} + (\frac{d S}{d x})^2
- \frac{d^2 S}{d x^2}]\]
\ho
= \frac{1}{2}R^{\dagger} R
\label{a1}
\la

\ho
R= i \frac{d}{d x} + i \frac{d S}{d x}
\label {a2}
\la

Then if we put $H^{FP}= H_{0} + U^{FP}(x) -E^{FP}$, we obtain the
Ricatti equation

\ho
U^{FP}(x) - E^{FP} =\frac{1}{2}[ (\frac{d S}{d x})^2 - \frac{d^2 S}{d x^2}]
\label{a3}
\la

Now suppose that $x\equiv \theta$ is compactified on a circle of radius 1.
Then it makes sense to work with the variables $Z= e^{i\theta}$ and rewrite
the wave function $\psi (\theta)$ as a single valued function of $Z$, i.e
$\psi (Z)$. Similarly, instead of working with the usual momentum
$p= -i \hbar \frac{d}{d \theta}$, we may choose the operator $X= e^{i p}$.
Both $X$ and $Z$ satisfy commutation relations

\ho
X Z = q Z X \,\,\,\,\,\,\,\,\,,\,\,\,\,\,\,\,\,\,q = e^{i \hbar}
\label{a4}
\la

Now we define the so-called $q$-derivative $D_{q}$

\ho
D_{q}= i \frac{X - X^{-1}}{q - q^{-1}}
\label{a5}
\la

\noindent which coincides with the usual $\frac{d}{d x}$ in the
limit $\hbar \longrightarrow 0$. The free Hamiltonian
$H_{0}= - \frac{1}{2}\frac{d^2}{d {\theta}^2}$ can therefore be replaced
by a ``$q$-deformed version'':

\ho
H_{0}(q)= - \frac{1}{2} D_{q}^{2}= \frac{1}{2(q - q^{-1})^2}(X^2 +X^{-2} - 2)
\label{a6}
\la

\noindent and the corresponding $FP$-Hamiltonian becomes:

\[H^{FP}(q)= \frac{1}{2(q - q^{-1})^2}[X^2 +X^{-2} - e^{S(Z)- S(q^{2} Z)} -
e^{S(Z)- S(q^{-2} Z)}]\]
\ho
=\frac{1}{2}R^{\dagger}(q) R(q)
\label{a7}
\la

\noindent with

\ho
R(q)= \frac{i}{q - q^{-1}}[e^{\frac{1}{2}[S(q^{2} Z) - S(Z)]}X^{2}-
e^{- \frac{1}{2}[S(q^{2} Z) - S(Z)]}]
\label{a8}
\la

\noindent where the hermiticity of $S$ has implicitely been assumed.
\vspace{0.3cm}

Finally, if we write $H^{FP}(q)$ as $H_{0}(q) + U^{FP}(Z) -E^{FP}$, we get

\ho
U^{FP}(Z) - E^{FP} = \frac{1}{2(q - q^{-1})^2}[2 - e^{S(Z)- S(q^{2} Z)} -
e^{S(Z)- S(q^{-2} Z)}]
\label{a9}
\la

Notice that this Ricatti equation determines the value of $S$ at the point
$q^2 Z$ as long as the values of $S$ at the points $Z$ and $q^{-2} Z$
are known. In this sense, (\ref{a9}) should  be considered as a second order
finite difference equation. Besides, either (\ref{a7}), (\ref{a8}) and
(\ref{a9}) become (\ref{a1}), (\ref{a2}) and (\ref{a3}), respectively,
in the ``classical''
limit $q \longrightarrow 1$. Moreover, if $q$ were a $k$-th root of
unity, then the
$\theta$ variable could be restricted to take a discrete set of values,
let us say
$\theta = \frac{2\pi n}{k}\,\,\,\,,\,\,\,\,n=0,1,\cdots,k-1$,
in which case we would obtain a ${\bf Z}_{k}$-invariant theory.

\appendix
\setcounter{section}{1}

\Section{ Exact FP-computation of D=1 finite chains}

A suggestive feature of the FP-approach considered in this paper is that
it provides a recipe to compute exactly the ground state and energy of a whole
family of $1$-dimensional hamiltonians defined on a finite lattice,
among which the ITF model of section 2 would constitute a particular
case. Here we present
the derivation of this result.

Firstly, let us consider the lattice configuration as a collection of
$0$'s and $1$'s. The presence
of a $1$ (respec. a $0$) at the site $j$ of the lattice should be thought of as
equivalent to the existence (respec. absence) of a $\sigma ^{z}$ at
this position.
Besides, let us assume that the number $N$ of sites of the lattice and
the number of $1$'s arranged on it are both  even.
Under these assumptions, it is not difficult to
convince oneself that there exist just $2^{N-1}$ different possibilities
to distribute a set of $0$'s and $1$'s on the lattice. We can associate
to each one of these possibilities a label $I_{a}$, namely:

\ho
\{ I_{a}{\}}_{a=1}^{2^{N-1}} \equiv \{ I_{1}=(0,0,\stackrel{(N)}{\cdots},0),
 I_{2}=(1,1,0\stackrel{(N)}{\cdots},0),
\cdots, I_{2^{N-1}}=(1,1,\stackrel{(N)}{\cdots},1,1)\}
\label{1b}
\la

Indeed, we can think in the labels $I_{a}$ as the elements of a
certain equivalence
class $\frac{{\bf Z}_{2}^{\otimes N}}{{\cal F}_{a}}$. Here ${\bf Z}_{2}$
 reflects the sole presence of $0$'s and $1$'s in the lattice, with the usual
 rule of sum:

\[1+0=1\,\,{\rm mod}\,\, 2\,\,\,\,\,\,\,\,,\,\,\,\,\,\,\,\,
1+1=0\,\,{\rm mod}\,\, 2\]

\noindent whereas ${\cal F}_{a}$ takes into account the fact that
 the number of $1$ is constrained to be  even:

\[{\cal F}_{a}=\sum_{j=1}^{N} I_{a}(j)=0\,\,{\rm mod}\,\, 2\]

Secondly, to each label $I_{a}$ we attach an operator
$\sigma_{I_{a}}^{z}$ defined as follows:

\ho
\sigma_{I_{a}}^{z}= \prod_{j=1}^{N} (\sigma_{j}^{z})^{I_{a}(j)}
\label{2b}
\la

\noindent ( to $I_{a}=(1,1,0,0,1,1)$ it would correspond the operator
$\sigma_{I_{a}}^{z}= \sigma_{1}^{z}\sigma_{2}^{z}
\sigma_{5}^{z}\sigma_{6}^{z}$, for example).

It is easy to see that the local operators $\sigma_{I_{a}}^{z}$
verify the fusion rule:

\ho
\sigma_{I_{a}}^{z} \sigma_{I_{b}}^{z}= \sigma_{I_{a}+I_{b}}^{z}
\label{3b}
\la

\noindent and, of course:

\ho
(\sigma_{I_{a}}^{z})^{2}={\bf 1}
\label{4b}
\la

Likewise, the following commutation relation is satisfied:

\ho
\sigma_{j}^{x} \sigma_{I_{a}}^{z} = (-1)^{I_{a}(j)}
\sigma_{I_{a}}^{z}\sigma_{j}^{x}
\label{5b}
\la

Finally, let us introduce a kind of generalized ITF hamiltonian,
characterized by the fact that the interacting term includes
couplings to all scales and to any number of spins:

\ho
H= - \sum_{j=1}^{N} \sigma_{j}^{x} - \sum_{a=2}^{2^{N-1}}
\lambda_{I_{a}} \sigma_{I_{a}}^{z}
\label{6b}
\la

The question of finding the ground state and energy of the
hamiltonian (\ref{6b})
can be faced as follows. Suppose we assume for the first the ansatz:

\ho
\mid \psi_{0} \rangle = e^{-\frac{1}{2} S(\sigma_{1}^{z},\cdots,
\sigma_{N}^{z}) }\mid 0 \rangle
\label{7b}
\la

\noindent but now with:

\ho
S(\sigma_{1}^{z},\cdots,\sigma_{N}^{z}) = - \sum_{a=2}^{2^{N-1}}
\alpha_{I_{a}} \sigma_{I_{a}}^{z}
\label{8b}
\la

\noindent and where the state $\mid 0 \rangle$ verify (\ref{34bis}).

We want to obtain the  FP hamiltonian associated to the ansatz
(\ref{7b}), so we
apply $H_{0}= - \sum_{j=1}^{N} \sigma_{j}^{x}$ on it, as to get:

\ho
H^{FP}= - \sum_{j=1}^{N} (\sigma_{j}^{x} - e^{\Delta_{j} (S)})= -
\sum_{j=1}^{N} (\sigma_{j}^{x} - e^{-     \sum_{a=2}^{2^{N-1}}
I_{a}(j)\alpha_{I_{a}} \sigma_{I_{a}}^{z}})
\label{9b}
\la

A lengthy but straightforward calculation, where the use of
eqs.(\ref{2b})-(\ref{5b})
is essential, allows one to rewrite (\ref{9b}) as:

\ho
H^{FP}=-\sum_{j=1}^{N}[\sigma_{j}^{x} - A_{j}\sum_{a=2}^{2^{N-1}}
\sigma_{I_{a}}^{z} \{ \sum_{p\geq 1}\frac{(-1)^{p}}{p !}
\sum_{\begin{array}{c}
a_{1},\cdots,a_{p}=2\\I_{a_{1}}+\cdots+ I_{a_{p}}=I_{a}
\end{array}}^ {2^{N-1}}  t_{I_{a_{1}}}^{(j)}\cdots t_{I_{a_{p}}}^{(j)}     \}]
\label{10b}
\la

\noindent where we have introduced

\ho
A_{j}= \prod_{a=2}^{2^{N-1}}\cosh [I_{a}(j)\alpha_{I_{a}}]
\label{11b}
\la

\ho
t_{I_{a}}^{(j)}\equiv I_{a}(j) t_{I_{a}}\equiv I_{a}(j) \tanh \alpha_{I_{a}}
\label{12b}
\la

In fact, if we take translational invariance for granted, the value of $A_{j}$
remains invariable whatever be the site $j$ on the lattice; we may then
express $A_{j} \equiv A,\,\, \forall j=1,\cdots,N$. Assuming that this
condition is fulfilled, the energy
associated with (\ref{10b}) is given by:

\ho
E^{FP}= - A [ N + \sum_{j=1}^{N} \sum_{p\geq 1} \frac{(-1)^{p}}{p !}
\sum_{\begin{array}{c}
a_{1},\cdots,a_{p}=2\\I_{a_{1}}+\cdots+ I_{a_{p}}=I_{1}
\end{array}}^ {2^{N-1}}
t_{I_{a_{1}}}^{(j)}\cdots t_{I_{a_{p}}}^{(j)} ]
\label{13b}
\la

Moreover,  the comparison between the $H^{FP}$ of (\ref{10b}) and the
generic hamiltonian (\ref{6b}) suggests the identification:

\ho
\lambda_{I_{a}}= - A \sum_{j=1}^{N} \sum_{p\geq1}
\frac{(-1)^{p}}{p !}
\sum_{\begin{array}{c}
a_{1},\cdots,a_{p}=2\\I_{a_{1}}+\cdots+ I_{a_{p}}=I_{a}
\end{array}}^ {2^{N-1}}
t_{I_{a_{1}}}^{(j)}\cdots t_{I_{a_{p}}}^{(j)} \,\,\,\,\,,\,\,\,\,\,
a=2,\cdots,{2^{N-1}}
\label{14b}
\la

\vspace{1cm}

To make explicit these formulae, let us apply them to the ITF model
of section 2, but now for the case of a finite lattice with $N=4$ sites.
In  table 1, we have summarized the notations and coupling
constants appearing in this case, as well as the interaction terms these
constants give account of.

A glance to table 1 suggests  the choice

\ho
\mid \psi_{0} \rangle = e^{    \frac{1}{2}\alpha_{1}\sum_{j=1}^{4}
\sigma_{j}^{z}
\sigma_{j+1}^{z} +  \frac{1}{4}\alpha_{2}\sum_{j=1}^{4}\sigma_{j}^{z}
\sigma_{j+2}^{z}+ \frac{1}{2}\alpha_{3}\sigma_{1}^{z}\sigma_{2}^{z}
\sigma_{3}^{z}\sigma_{4}^{z}} \mid 0 \rangle
\label{15b}
\la

\noindent where the factor
$\frac{1}{4}$ in (\ref{15b})  has been introduced in order to avoid
overcounting in the corresponding term (for $N=4$, \,\,\,$\sum_{j=1}^{4}
\sigma_{j}^{z}
\sigma_{j+1}^{z}=\sum_{j=1}^{4}\sigma_{j}^{z}
\sigma_{j+3}^{z}$).

 Now we  require that $H^{FP}_{ITF}$ coincides with $H_{ITF}$:

\ho
H^{FP}_{ITF}=H_{ITF}= -\sum_{j=1}^{4}\sigma_{j}^{x} - \lambda \sum_{j=1}^{4}
\sigma_{j}^{z}\sigma_{j+1}^{z}
\label{16b}
\la

\noindent but since we have three different couplings  it must be

\[t_{I_{2}}=t_{I_{3}}=t_{I_{4}}=t_{I_{5}}\equiv t_{1}=\tanh \alpha_{1}\]
\ho
t_{I_{6}}=t_{I_{7}}\equiv t_{2}=\tanh \alpha_{2}
\label{17b}
\la
\[t_{I_{8}}\equiv t_{3}=\tanh \alpha_{3}\]

On the other hand, one is free to select a particular site, say
$j = 1$, to get:

\ho
A=A_{1}=\prod_{a=2}^{8} \cosh [I_{a}(1) \alpha_{I_{a}}]= \frac{1}
{(1-t_{1}^{2}) (1-t_{2}^{2})^{1/2} (1-t_{3}^{2})^{1/2}}
\label{18b}
\la

Now, introducing eqs.(\ref{17b})-(\ref{18b}) in (\ref{13b})-(\ref{14b})
we obtain:

\ho
e^{FP}_{ITF}=-\frac{E^{FP}_{ITF}}{4}=  \frac{1 + t_{1}^{2} t_{2} t_{3}}
{ (1-t_{1}^{2}) (1-t_{2}^{2})^{1/2} (1-t_{3}^{2})^{1/2}}
\label{19b}
\la

\noindent and

\[\lambda_{I_{2}}=\lambda_{I_{3}}=\lambda_{I_{4}}=\lambda_{I_{5}}
=\lambda= \frac{2t_{1}(1-t_{2})(1-t_{3})}
{ (1-t_{1}^{2}) (1-t_{2}^{2})^{1/2} (1-t_{3}^{2})^{1/2}}\]
\ho
\lambda_{I_{6}}=\lambda_{I_{7}}=0=(1-t_{3})(t_{1}^{2}-t_{2})
\label{20b}
\la
\[\lambda_{I_{8}}=0= t_{3} + t_{1}^{2} t_{2}\]

Eqs.(\ref{19b})-(\ref{20b}) can be easily solved, and in fact we obtain:

\ho
e^{FP}_{ITF}=\frac{(1+t_{2}^{2})^{1/2}}{1-t_{2}}
\label{21b}
\la

\noindent where $t_{2}$ is the real and positive solution of the polinomial:

\ho
t_{2}^{4} -\frac{4t_{2}^{3}}{\lambda^{2}} -2t_{2}^{2}
-\frac{4 t_{2}}{\lambda^{2}} + 1 =0
\label{22b}
\la

\noindent namely:

\ho
t_{2}(\lambda)= \frac{1}{\lambda^{2} -1 + \sqrt{1 + \lambda^{4}}}
[\lambda^{2} +1+\sqrt{1 + \lambda^{4}} -2 (\lambda^{2}+ \sqrt{1 +
\lambda^{4}})^{1/2}]
\label{23b}
\la

Now if we substitute eq.(\ref{23b}) in (\ref{21b}) we find the final result:

\ho
e^{FP}_{ITF}(\lambda)= \frac{1}{2}[\sqrt{1 + \lambda^{2} + \lambda \sqrt{2}} +
\sqrt{1 + \lambda^{2} - \lambda \sqrt{2}}]
\label{24b}
\la

This expression  coincides  precisely with the exact value given by the
formula \cite{Pfeuty1}:

\ho
e_{ITF}(\lambda)= \frac{1}{N} \sum_{j=-\frac{N}{2}}^{\frac{N}{2}-1}
\sqrt{  1 + \lambda^{2} +2\lambda \cos [ \frac{(2j +1)\pi}{N}]}
\label{25b}
\la

\noindent when we particularized to the lattice with $N=4$.

\appendix
\setcounter{section}{2}

\Section{ Explicit computations in the FP-version of the ITF model}

In this appendix we deduce the formulae for the $t_{n}$ and $e_{N}^{FP}$
exhibited in section 4.

The case $u_{N}=1$ is very easy to handle because the square roots appearing in
eqs.(\ref{59})-(\ref{59bis}) can be performed in terms of trigonometric
functions. In particular, we get from eq.(\ref{56})

\ho
{\hat t}_{m} = -\sqrt{\frac{2C_{N}}{N}} \,\,\sin\,(\frac{\pi m}{N})
\label{1c}
\la

\noindent and introducing this equation into eq.(\ref{57}) we find:

\ho
1 =
\frac{\sqrt{2 C_{N}}}{N} \sum_{m=0}^{N-1} \sin \,(\frac{\pi m}{N})
=  \frac{\sqrt{2 C_{N}}}{N} \cot \,(\frac{\pi }{2N})
\label{2c}
\la

{}From this euation the value for $C_{N}$ follows:

\ho
C_{N} = \frac{1}{2} (N \tan \,\frac{\pi }{2N})^{2}
\label{61}
\la

On the other hand, working out eq.(\ref{59}) with the help of
(\ref{61}) and (\ref{1c}) yields

\ho
t_{n}(N) = \frac{{\sin}^{2} \,(\frac{\pi }{2N})}
{\sin \,[\frac{\pi}{N}\, (n + \frac{1}{2})]\,\,
\sin \,[\frac{\pi}{N}\, (n + \frac{1}{2})]}
\label{c4}
\la

In order to compute the energy $e^{FP}_{N}$ we rewrite eq.(\ref{52})
in the form

\ho
e^{FP}_{N} = \prod _{n=1}^{N-1} (1 - t_{n} )^{-\frac{1}{2}}\,\,
(1 + t_{n} )^{-\frac{1}{2}}
\label{c5}
\la

Now we can make use of eq.(\ref{c4}) to express the monomials of (\ref{c5})
as
\ho
1 + t_{n} = \frac{{\sin}^{2}\, \,(\frac{\pi n}{N})}
{\sin \,[\frac{\pi}{N}\, (n + \frac{1}{2})]\,\,
\sin \,[\frac{\pi}{N}\, (n + \frac{1}{2})]}
\label{c6}
\la

\ho
1 - t_{n} = \frac{ \sin\, \,(\frac{\pi n}{N} + \frac{\theta_{N}}{2})
 \sin\, \,(\frac{\pi n}{N} - \frac{\theta_{N}}{2})}
{\sin \,[\frac{\pi}{N}\, (n + \frac{1}{2})]\,\,
\sin \,[\frac{\pi}{N}\, (n + \frac{1}{2})]}
\label{c7}
\la

\noindent where $\theta_{N}$ was defined in equation (\ref{62penta}).

Now using the formula

\ho
\prod_{n=1}^{N-1} \sin \,\frac{\pi}{N} \,(n - a) = \frac{1}{2^{N-1}}
\,\,\frac{\sin (\pi a)}{\sin (\frac{\pi a }{N})}
\label{c9}
\la

\noindent we obtain

\ho
\prod_{n=1}^{N-1} ( 1 + t_{n}) = (N \sin \,\,\frac{\pi}{2N})^{2}
\label{c10}
\la

\ho
\prod_{n=1}^{N-1} ( 1 - t_{n}) = \left( \frac{\sin \frac{\pi}{2N}\,\,
 \sin \frac{N \theta_{N}}{2}}{\sin \frac{\theta_{N}}{2}}\right)^{2}
\label{c11}
\la

{}From these equations and eq.(\ref{63}) the eq.(\ref{61}) follows directly.

\vspace{0.8cm}
Next, let us consider the case where $N \longrightarrow \infty$ and
$ 0 \leq u_{\infty} \leq 1$. Now the  sums can be converted into integrals,
as for example eq.(\ref{59}):

\[
t_{n}= - \int_{0}^{2 \pi} \,\frac{dx}{2 \pi} \,e^{i n x} (C_{\infty} -
 \frac{2 \lambda}{e_{\infty}^{FP}} \cos x)^{1/2}\]
\ho
= (-1)^{n + 1} \frac{\sqrt{C_{\infty}}}{\pi} \int_{0}^{\pi} dx \,\cos(nx)
( 1 + u_{\infty} \cos x)^{1/2}
\label{c13}
\la

 But replacing \,\,\, $\cos x = 1 - 2 {\sin}^{2} (\frac{x}{2})$ \,\,in
(\ref{c13}) and subsequently changing
variables $\frac{x}{2} \longrightarrow x$ we get:

\ho
t_{n} =  (-1)^{n + 1}\frac{2}{\pi} \sqrt{C_{\infty}(1 + u_{\infty})}
\int_{0}^{\frac{\pi}{2}} \cos(2nx)\,\,(1 - \frac{2u_{\infty}}{1
+ u_{\infty}} {\sin}^{2} x )^{1/2}
\label{c14}
\la

This integral can be performed if we first expand the square root
according to the formula:

\ho
(1- Z)^{\alpha} = \sum_{m=0}^{\infty} \frac{\Gamma (m - \alpha)}{
\Gamma(-\alpha)\Gamma (m + 1)}\,\,
\,\,Z^{m}
\label{c15}
\la

\noindent and then we use the result

\ho
\int_{0}^{\frac{\pi}{2}} dx \,\,\cos(2nx) \,\,{\sin}^{2m} x =
\left\{ \begin{array}{ll}
\frac{(-1)^{n}}{2^{2m}} \frac{\Gamma (2m + 1)}{\Gamma(m + n + 1)\,\,
\Gamma(m - n + 1)}& m\geq n\\
0&m < n \end{array}
\right.
\label{c16}
\la

Finally, we obtain the expression

\ho
t_{n} =\frac{ \sqrt{2 C_{\infty}}}{8 \pi} \frac{u_{\infty}^{n}}
{(1 + u _{\infty})^{n-\frac{1}{2}}}\,\,
\frac{\Gamma(\frac{n}{2} - \frac{1}{4})\,\,\Gamma(\frac{n}{2} + \frac{1}{4})}
{\Gamma(n + 1)}\,\, F(n - \frac{1}{2}, n + \frac{1}{2}, 2n + 1; \frac{2
u_{\infty}}{1 + u_{\infty}} )
\label{c17}
\la

The equation which yields $C_{\infty}$ is obtained by setting $t_{0}=-1$ in
eq.(\ref{c17}), namely

\ho
\sqrt{C_{\infty}} ( 1 + u_{\infty})^{1/2} \,\,F(-\frac{1}{2}, \frac{1}{2},1;
\frac{2 u_{\infty}}
{1 + u_{\infty}} ) = 1
\label{c18}
\la

Taking the square of eq.(\ref{c18}) and using the fact that $C_{\infty}=
\frac{2 \lambda}{e_{\infty}^{FP} u_{\infty}}$, we deduce an equation for
$e_{\infty}^{FP}$as a function of $\lambda$ and $u_{\infty}$:

\ho
e_{\infty}^{FP} = 2 \lambda (\frac {1 + u_{\infty}}{u_{\infty}}) \,\,
F^{2}(-\frac{1}{2}, \frac{1}{2},1;
\frac{2 u_{\infty}}
{1 + u_{\infty}} )
\label{c19}
\la

Replacing now $u_{\infty}$ in terms of the parameter $\lambda_{R}$
introduced in (\ref{67p}) one obtains eq.(\ref{66p}).

As for the equation (\ref{65p}) for the $t_{n}$'s, it is easily
derived from (\ref{67p}) and
(\ref{c17})-(\ref{c18}) . Likewise, expression (\ref{66p}) is obtained
by taking the $N\longrightarrow \infty$ limit of (\ref{c5}):

\ho
e^{FP}_{\infty} = \prod _{n=1}^{\infty} (1 - t_{n}^{2} )^{-1}
\label{c20}
\la

\noindent and using  (\ref{c17}). Notice that the exponent in
eq.(\ref{c20}) is $-1$
and not $-\frac{1}{2}$ as in eq.(\ref{c5}). This takes care of the constraint
$t_{n}= t_{N-n}$ which in the $N\longrightarrow \infty$ limit becomes
$t_{n} = t_{-n}$, with $n\in {\bf Z}$.

Eq. (\ref{c17}) for $u_{\infty}=1$ gives the value
$t_{n}= \frac{1}{4 n^{2} -1}$. If
$u_{\infty}<1$ we can also find an indication of the behaviour of the
$t_{n}$, just on
condition that $n$ be large enough. To this end, we can carry out a
saddle point evaluation
for $t_{n}$ on the basis of the standard integral
 representation of hypergeometric functions \cite{Erdel}.
Indeed, we can write (\ref{c17}) as:

\ho
t_{n} = \frac{\sqrt{C_{\infty} ( 1 + u_{\infty})}}{2 \pi} \,\,
\frac{z^{n}}{(n- \frac{1}{2})}
\,\,\int_{0}^{1} dt\, \left[ \frac{t(1-t)}{1 - zt} \right]^{n- \frac{1}{2}}
\label{c21}
\la

\noindent where we have introduced the parameter

\ho
z=\frac{2 u_{\infty}}{1 + u_{\infty}}
\label{c21bis}
\la

Due to the particularly compact form of the integrand, the saddle point in
eq.(\ref{c21}) turns out to be independent of $n$. Our result is

\[
t_{n} \simeq \frac{1}{2} \sqrt{\frac{C_{\infty}}{\pi}}\,\,
\frac{1}{(n - \frac{1}{2})^{\frac{3}{2}}} \]
\ho
\times \left(
\frac {1 + \sqrt{1 - u_{\infty}^{2}}}{1 + u_{\infty} + \frac{1}{2}
\sqrt{1 - u_{\infty}^{2}}}\right)^{\frac{1}{2}}\,\,\left(\frac{1
- u_{\infty}}{1 + u_{\infty}}\right)^{\frac{1}{4}}\,\,
 \left(\frac{1 + \sqrt{1 - u_{\infty}^{2}}}{u_{\infty}}\right)^{-n}
\,\,\,\,\,\,\,n>>1
\label{c22}
\la

Finally, the expression (\ref{67p}) can be used to replace
$u_{\infty}$ by $\lambda_{R}$
and in this way we get eq.(\ref{71p}).

\appendix
\setcounter{section}{3}

\Section{ Higher dimensional ansatzs}

One of the reasons to justify the introduction of new ansatzs
 for the ground state of the $1D$ ITF model, amounts to
our wish of discussing higher dimensional models
 for which exact solutions are not known. In this sense, the obvious
generalization of the $1D$-dimensional ansatz proposed in eq.(\ref{48}) becomes

\ho
\mid \psi \,(\{\alpha_{{\bf r}}\})\rangle = \exp \left(\frac{1}{4}
\sum_{{\bf r}\in \Lambda'}
\sum_{{\bf j}\in \Lambda} \alpha_{{\bf r}} \,\,\sigma_{{\bf j}}^{z}
\sigma_{{\bf j} + {\bf r}}^{z}
\right) \,\mid 0 \rangle
\label{d1}
\la

\noindent where $\Lambda$ is the hypercubic lattice ${\bf Z}_{L}^{D}$ and
$\Lambda' = \Lambda  -\{0\}$. We shall suppose that $\alpha_{{\bf r}} =
\alpha_{-{\bf r}}$
with ${\bf r} \in \Lambda'$.

The FP-Hamiltonian corresponding to (\ref{d1}) has the form

\ho
H^{FP}= - \sum_{{\bf j}\in \Lambda}[ \sigma_{{\bf j}}^{x} - e^{
- \sum_{{\bf r}\in \Lambda'}^{}
 \alpha_{{\bf r}} \,\,\sigma_{{\bf j}}^{z} \sigma_{{\bf j} + {\bf r}}^{z}} ]
\label{d2}
\la

Expanding eq.(\ref{d2}) we arrive to a generalized ITF model

\ho
H^{FP} = - E^{FP} - \sum_{{\bf j}\in \Lambda} \sigma_{{\bf j}}^{x} -
\frac{1}{2} \sum_{{\bf j}\in \Lambda}
 \lambda_{{\bf r}} \,\,\sigma_{{\bf j}}^{z} \sigma_{{\bf j} + {\bf r}}^{z} +
( 4 \sigma^{z}{\rm 's}) + \cdots
\label{d3}
\la

\noindent with

\ho
E^{FP}= - N e_{N}^{FP}= -N \prod_{{\bf r}\in \Lambda '} \cosh \alpha_{{\bf r}}
\label{d4}
\la

\ho
\frac{\lambda_{{\bf r} }}{e_{N}^{FP}}=-
\sum_{{\bf n}\in \Lambda}
t_{{\bf n}}t_{{\bf r} - {\bf n}}
\label{d5}
\la

\noindent and

\[ t_{{\bf 0}}= -1 \]
\ho
t_{{\bf r}}= \tanh \alpha_{{\bf r}} \,\,\,\,\,\,,\,\,\,\,\,\,
{\bf r}\neq {\bf 0}
\label{d6}
\la

To solve eqs.(\ref{d4})-(\ref{d5}) we introduce the Fourier transformation

\ho
t_{{\bf n}} = \frac{1}{L^{D/2}} \sum_{{\bf m} \in \Lambda^{\ast}}
e^{2 \pi i \,\frac{{\bf n}
{\bf m}}{L}} \,{\hat t}_{\bf m}
\label{d7}
\la

\noindent where $\Lambda^{\ast}$ is the dual lattice of $\Lambda$.
For a hypercubic lattice
both $\Lambda^{\ast}$ and $\Lambda$ coincide. Following the same steps
of section 4, we find

\ho
{\hat t}_{\bf m} = - \frac{\varepsilon_{\bf m}}{L^{D/2}} \left(
C_{N} - \sum_{{\bf r} \in \Lambda '} e^{- 2 \pi i
\,\frac{{\bf r}
{\bf m}}{L}} \,
\frac{\lambda_{{\bf r} }}{e_{N}^{FP}}\right)^{1/2}
\label{d8}
\la

\noindent where

\ho
C_{N}= \sum_{{\bf m} \in \Lambda^{\ast}} {\hat t}^{2}_{{\bf m}} =
\sum_{{\bf n} \in \Lambda} t^{2}_{{\bf n}}
\label{d9}
\la

\noindent and $\varepsilon_{{\bf m}} = \varepsilon_{- {\bf m}}= \pm 1$.
For convenience,
we shall fix $\varepsilon_{{\bf m}} = 1$.

In the limit $N\longrightarrow$ the $t_{{\bf n}}$ are given by

\ho
t_{{\bf n}} = - \int_{0}^{2 \pi} \frac{d k_{1}}{2 \pi}\cdots \int
_{0}^{2 \pi} \frac{d k_{D}}{2 \pi}\,\,e^{i {\bf n} {\bf k}}
\,\,\left[ C_{\infty} - \frac{2}{e^{FP}_{\infty}}
\sum_{a=1}^{D} \lambda_{a} \,\cos k_{a} \right]^{1/2}
\label{d10}
\la

\noindent where $\lambda_{a}$ is the value of the coupling constant multiplying
the operator $\sigma_{{\bf j}}^{z} \sigma_{{\bf j} + \mu_{a}}^{z}$ in
eq.(\ref{d3}) ($\mu_{a}$ is the unit vector along the $a^{th}$-direction).

In the special case of $D=2$ the integrals of (\ref{d10}) can be explicitly
determined in a way which goes parallel to the followed in appendix C.~Writting
${\bf n} =(n_{1},n_{2})$  and making use of eq.(\ref{d6}) we obtain
after a lengthy computation
\vspace{0.3cm}

\[
e^{FP}_{\infty} = \frac{2 (\lambda_{1} + \lambda_{2})}
{\pi^{5} (\lambda_{R_{1}} + \lambda_{R_{2}})}\]
\ho
\times \left[ ( 1 + \lambda_{R_{1}} + \lambda_{R_{2}})\,\,
 F_{2} (- \frac{1}{2}, \frac{1}{2},
\frac{1}{2} ,  1,1 ,  \frac{4 \lambda_{R_{1}}}{(1 +
\lambda_{R_{1}} + \lambda_{R_{2}})^{2}},\,\, \frac{4 \lambda_{R_{2}}}{(1 +
\lambda_{R_{1}} + \lambda_{R_{2}})^{2}}) \right]^{2}
\label{d10bis}
\la
\vspace{0.3cm}

\noindent and\\
$\phantom{x}$\newpage

\[
t_{n_{1},n_{2}} = \frac{1}{2 \pi \sqrt{\pi}}  \]
\[\times \left[ \frac{4 \lambda_{R_{1}}}{(1 +
\lambda_{R_{1}} + \lambda_{R_{2}})^{2}}\right]^{n_{1}}
\,\, \left[ \frac{4 \lambda_{R_{2}}}{(1 +
\lambda_{R_{1}} + \lambda_{R_{2}})^{2}}\right]^{n_{2}} \,\,
\frac{\Gamma(n_{1} + n_{2} - \frac{1}{2}) \, \Gamma (n_{1} + \frac{1}{2}) \,
\Gamma (n_{2} + \frac{1}{2})}{\Gamma (2 n_{1} + 1) \,\Gamma (2 n_{2} + 1)}\]
\ho
\times \frac{F_{2} (n_{1} + n_{2}- \frac{1}{2}, n_{1}+ \frac{1}{2},
n_{2} + \frac{1}{2} , 2 n_{1} + 1,2 n_{2} + 1 ,  \frac{4 \lambda_{R_{1}}}{(1 +
\lambda_{R_{1}} + \lambda_{R_{2}})^{2}}, \,\,\frac{4 \lambda_{R_{2}}}{(1 +
\lambda_{R_{1}} + \lambda_{R_{2}})^{2}})     }
{  F_{2} (- \frac{1}{2}, \frac{1}{2},
\frac{1}{2} ,  1,1 ,  \frac{4 \lambda_{R_{1}}}{(1 +
\lambda_{R_{1}} + \lambda_{R_{2}})^{2}},\,\, \frac{4 \lambda_{R_{2}}}{(1 +
\lambda_{R_{1}} + \lambda_{R_{2}})^{2}})    }
\label{d10tris}
\la

\noindent where by $F_{2}$ we denote one of the hypergeometric series in
two variables
\cite{Erdel}. The parameters $\lambda_{R_{a}}\,\,,a=1,2$  play the role of
``dressed'' coupling constants and are defined through

\ho
\frac{\lambda_{a}}{C_{\infty} e^{FP}_{\infty}} = \frac{\lambda_{R_{a}}}
{1 + (\lambda_{R_{1}} + \lambda_{R_{2}})^{2}  }\,\,\,\,\,\,,\,\,\,\,\,\,a=1,2
\label{d10tetra}
\la

{}From these formulae it is possible to investigate the existence of
critical behaviour in
the same way as we did in section 4 for the one dimensional case. Nevertheless,
this analysis can be carried out directly from
 eq.(\ref{d10}). To this respect,
 we impose the condition

\ho
C_{\infty} = \frac{2}{e^{FP}_{\infty}} \sum_{a=1}^{D} \lambda_{a}
\label{d11}
\la

For large values of $\mid {\bf n} \mid$ the integral (\ref{d10}) is
dominated by the small momenta
$\mid {\bf k} \mid < \epsilon$, where $\epsilon \sim \frac{1}{\mid
{\bf n} \mid}$ is a cutoff; namely

\ho
t_{{\bf n}} \simeq - \frac{1}{e^{FP}_{\infty}} \,\int_{\mid {\bf k}
\mid < \epsilon}
\prod_{a=1}^{D} \frac{d k_{a}}{2 \pi}\,\,e^{i {\bf n} {\bf k}}\,\,
(\sum_{a=1}^{D} \lambda_{a} k_{a}^{2} )^{1/2}
\label{d12}
\la

\noindent and assuming isotropy, i.e  $\lambda_{a}=\lambda,\,\,\,
\forall a$, we obtain the scaling
behaviour

\ho
t_{{\bf n}} \sim \frac{1} {\mid {\bf n} \mid ^{D +1}}
\label{d13}
\la

Away from the condition (\ref{d11}) we shall define a ``mass parameter''
$\mu^{2}$ as

\ho
\mu^{2}= \frac{C_{\infty} e_{\infty}^{FP}}{\lambda} - 2D
\label{d14}
\la

Because the isotropy, the relation between $\mu$ and the sole
``dressed'' coupling constant $\lambda_{R}$ existing in this case, comes from
(\ref{d10tetra}) and is given by

\ho
\mu = \frac{1 - D\lambda_{R}}{\sqrt{\lambda_{R}}}
\label{d14bis}
\la

In terms of $\mu$, the behaviour of the $t_{{\bf n}} $ for large values
of $\mid {\bf n} \mid$ is dictated by

\ho
t_{{\bf n}} \simeq - \sqrt{\frac{\lambda}{e^{FP}_{\infty}}} \,
\int_{\mid {\bf k} \mid < \epsilon}
\prod_{a=1}^{D} \frac{d k_{a}}{2 \pi}\,\,e^{i {\bf n} {\bf k}}\,\,(
\mu^{2} + \mid {\bf k} \mid  ^{2} )^{1/2}
\label{d15}
\la

Moreover, a saddle point evaluation of this integral for $\mu^{2} >0$ yields
the exponential decaying law

\ho
t_{{\bf n}} \sim \frac{1} {\mid {\bf n} \mid ^{\frac{D}{2} +1}}\,\,e^{-
\mu \mid {\bf n} \mid }
\label{d16}
\la

Hence, $1/\mu$ defines the correlation length.

The constraint $t_{{\bf 0}}= -1$ gives us an expresion where the
density of energy
$e^{FP}_{\infty}$ and the parameter $\mu$ are tied together, namely

\ho
\left( \frac{e^{FP}_{\infty}}{\lambda}\right)^{1/2} =
\int_{0}^{2 \pi} \frac{d k_{1}}{2 \pi}\cdots
\int_{0}^{2 \pi} \frac{d k_{D}}{2 \pi} \left[ \mu^{2} + 4 \sum_{a=1}^{D}
\sin^{2} (\frac{k_{a}}{2}) \right]^{1/2}
\label{d17}
\la

To investigate the singular behaviour of $e^{FP}_{\infty}$ as a
function of $\mu$, it is
sufficient to consider the small values of the ``momenta'' $k_{a}$ as we
did previously for the
$t_{{\bf n}}$. In this case the integral (\ref{d17}) becomes
proportional to the function $F_{D}(\mu, \epsilon)$ defined as

\ho
F_{D}(\mu, \epsilon) = \int_{0}^{\epsilon} d\,k \,\,k^{D-1} \,
(\mu^{2} + k^{2})^{1/2}
\label{d19}
\la

\noindent where $\epsilon$ is a cutoff. Studying the singular behaviour
of $F_{D}(\mu, \epsilon)$ in the vecinity of the point $\mu =0$ we get:

\ho
F_{D}(\mu, \epsilon) = \left\{
\begin{array}{ll}
c_{D}\,\, \mu^{D +1} \log (\frac{\mu} {\epsilon}) + F_{D}^{reg}(\mu,
 \epsilon)&\,\,\,\, {\rm for} \,\,\,D \,\,{\rm odd}\\
&\\
 F_{D}^{reg}(\mu,
 \epsilon)&\,\,\,\, {\rm for} \,\,\,D \,\,{\rm even}
\end{array}\right.
\label{d20}
\la

\noindent where $F_{D}^{reg}(\mu,\epsilon)$ denotes a regular function
at $\mu =0$. Eq. (\ref{d20}) means
that only in odd dimensions ($D=3,5,\cdots$) we obtain a singular
logarithmic behaviour; that
is somehow the generalization of the one dimensional result
(\ref{70p}). The only difference
with this latter is the power $\mu^{D+1}$ in front of the logarithm,
which is expected from dimensional grounds.

Curiously enough, we do not observe singularities in the ground state energy
if the dimension is even. This fact does not exclude however the existence
of discontinuities in some derivative of the energy as we cross the
critical point, just as it occurs in a mean field calculation.

In any case, it is clear that the ansatz (\ref{d1}) does not lead for $D>1$ to
the standard ITF model. That supposes, according to the general ideas
expressed in
the introduction of the paper, that we have to return to perturbation theory
and figure out the terms necessary to improve the ansatz. In $D=2$ for example,
we have neglected in the ansatz the terms involving four
${\sigma^{z}}$ matrices. We claim here that these
terms are important if one wants to get closer to the ITF  model.
As a result, we propose the following ansatz for $D\geq 2$

\ho
\mid \psi (\{\alpha_{{\bf r}_{1},{\bf r}_{2},{\bf r}_{3}}\}) \rangle =
\exp ( \sum_{{\bf r}{\rm 's}\in \Lambda'} \frac{\alpha_{{\bf r}_{1},
{\bf r}_{2},{\bf r}_{3}}}
{2 n_{{\bf r}_{1},{\bf r}_{2},{\bf r}_{3}}} \sum_{{\bf j} \in \Lambda}
\sigma_{{\bf j}}^{z}\sigma_{{\bf j}+ {\bf r}_{1}}^{z}\sigma_{{\bf j}
+ {\bf r}_{2}}^{z}\sigma_{{\bf j}+ {\bf r}_{3}}^{z})\mid 0 \rangle
\label{d21}
\la

\noindent where $n_{{\bf r}_{1},{\bf r}_{2},{\bf r}_{3}}$ is defined by

\ho
n_{{\bf r}_{1},{\bf r}_{2},{\bf r}_{3}}= \left\{
\begin{array}{ll}
2& {\rm if} \,\,\,{\bf r}_{a}={\bf r}_{b}\neq{\bf r}_{c}\\
4& {\rm otherwise}
\end{array}\right.
\label{d22}
\la

Equation (\ref{d21}) includes the ansatz (\ref{d1}) by allowing two
${\bf r}{\rm 's}$ to coincide.
The quantities (\ref{d22}) accounts for the overcounting in the sum
over ${\bf j}$ in the
exponent of (\ref{d21}). Likewise the parameters
$\alpha_{{\bf r}_{1},{\bf r}_{2},{\bf r}_{3}}$ has to be
totally symmetric in their three indices and must satisfy the
reflection symmetry property

\ho
\alpha_{{\bf r}_{1},{\bf r}_{2},{\bf r}_{3}}=
 \alpha_{-{\bf r}_{1},{\bf r}_{2}-{\bf r}_{1},{\bf r}_{3}-{\bf r}_{1}}=
\alpha_{{\bf r}_{1}-{\bf r}_{2},-{\bf r}_{2},{\bf r}_{3}-{\bf r}_{2}}
=\alpha_{{\bf r}_{1}-{\bf r}_{3},{\bf r}_{2}-{\bf r}_{3},-{\bf r}_{3}}
\label{d23}
\la

\noindent which is the analogue of condition $\alpha_{r}= \alpha_{-r}$
in the one dimensional case.

The FP-Hamiltonian associated to (\ref{d21}) is

\ho
H^{FP}= - \sum_{{\bf j} \in \Lambda} [\sigma_{{\bf j}}^{x} -
\exp(- \sum_{{\bf r}{\rm 's}\in \Lambda'}  \alpha_{{\bf r}_{1},
{\bf r}_{2},{\bf r}_{3}}
\sigma_{{\bf j}}^{z}\sigma_{{\bf j}+ {\bf r}_{1}}^{z}\sigma_{{\bf j}
+ {\bf r}_{2}}^{z}\sigma_{{\bf j}+ {\bf r}_{3}}^{z})]
\label{d24}
\la

\noindent and can be expanded as

\ho
H^{FP} = - E^{FP} - \sum_{{\bf j} \in \Lambda} \sigma_{{\bf j}}^{x}
- \sum_{{\bf r}{\rm 's}\in \Lambda'}
\frac{\lambda_{{\bf r}_{1},{\bf r}_{2},{\bf r}_{3}}}
{ (n_{{\bf r}_{1},{\bf r}_{2},{\bf r}_{3}})!}
\sum_{{\bf j} \in \Lambda}
\sigma_{{\bf j}}^{z}\sigma_{{\bf j}+ {\bf r}_{1}}^{z}\sigma_{{\bf j}
+ {\bf r}_{2}}^{z}\sigma_{{\bf j}+ {\bf r}_{3}}^{z} + ( 6
\sigma^{z}{\rm 's}) + \cdots
\label{d25}
\la

Compairing (\ref{d24}) and (\ref{d25}) and using the same trick of
section 4, according to
which one could replace consistently the $\sigma_{{\bf j}}^{z}$ by
$\sigma_{{\bf j}}= \pm1$,
we obtain the following equations  for $E^{FP}$ and
$\lambda_{{\bf r}_{1},{\bf r}_{2},{\bf r}_{3}}$ as functions of
$\alpha_{{\bf r}_{1},{\bf r}_{2},{\bf r}_{3}}$

\ho
E^{FP}= -\frac{N}{2^{N}} \sum_{\sigma{\rm 's}}^{} \exp(-\sum_{{
\bf r}{\rm 's}\in \Lambda'}
\alpha_{{\bf r}_{1},{\bf r}_{2},{\bf r}_{3}}
\sigma_{{\bf 0}}\sigma_{ {\bf r}_{1}}\sigma_{
 {\bf r}_{2}}\sigma_{ {\bf r}_{3}})
\label{d26}
\la

\ho
\lambda_{{\bf s}_{1},{\bf s}_{2},{\bf s}_{3}}= -\frac{1}{2^{N}}
\sum_{\sigma{\rm 's}}^{}(\sum_{{\bf j} \in \Lambda}
\sigma_{{\bf j}}\sigma_{{\bf j}+ {\bf s}_{1}}\sigma_{{\bf j}
+ {\bf s}_{2}}\sigma_{{\bf j}+ {\bf s}_{3}})
\exp(-\sum_{{\bf r}{\rm 's}\in \Lambda'}
\alpha_{{\bf r}_{1},{\bf r}_{2},{\bf r}_{3}}
\sigma_{{\bf 0}}\sigma_{ {\bf r}_{1}}\sigma_{
+ {\bf r}_{2}}\sigma_{ {\bf r}_{3}})
\label{d27}
\la

Naturally the problem which arises now is to perform the sum
over the $\sigma$'s.
Observe that eq.(\ref{d26}) is a kind of classical partition
function of a lattice
involving couplings between 2 and 4 spins; in some sense, it recalls an
8-vertex model \cite{Baxter} so that there are chances to perform
the sum, at least
in some particular cases.

Finally, there is the question of inverting eq.(\ref{d27}) in order to express
the $\alpha$'s as  functions of the $\lambda$'s.
Notice that the $\lambda$'s could be thought of as a kind of correlators
among $\sigma$'s for a partition function as the one of eq.(\ref{d26}).

\vspace{0.3cm}
In summary, the problem of constructing the ground state for the
$2D$ quantum ITF
model is essentially reduced, by means of the ansatz (\ref{d21}), to that of
computing a $2D$ nonlocal partition function followed by the inversion of
the relation between certain correlators and coupling constants.
The difficulties
of these two tasks are considerable but we expect that the FP procedure
 can throw some light on the original unsolved problem.

\newpage
\noindent {\Large{\bf Table captions}}
\vspace{1.5cm}

\noindent{\bf Table1}: Values of the coupling constants and interaction
terms in the case of
a ITF model defined on a finite lattice with $N=4$ sites.
\newpage
\noindent {\Large{\bf Figure captions}}
\vspace{1.5cm}

\noindent{\bf Figure 1a)}.- The shadow region simbolizes the possible
non vanishing entries of the matrix $p_{n,I}$.

\noindent{\bf Figure 1b)}.- Information obtained up to order $\nu$ in
perturbation theory.

\noindent{\bf Figure 1c)}.- Information obtained up to order $\nu$
using a truncated cluster expansion.

\noindent{\bf Figure 1d)}.- Here we choose one or more representants
for each order in perturbation
theory and construct a Hamiltonian which belongs to the same universality class
of the original Hamiltonian.

\noindent{\bf Figure 2)}.- Comparison of the values for minus the 1D energy
density in the cases
exact $e_{Ising,D=1}$ (strong solid line), in the
 first order FP-aproximation $e_{Ising,D=1}^{FP,(1)}$ (solid line) and in the
 second  order FP-aproximation $e_{Ising,D=1}^{FP,(2)}$ (grey line).

\noindent{\bf Figure 3)}.- Action of the operator $\Delta$ on a generic state
$\mid I_{a} \rangle = \mid \mu_{1},\cdots,\mu_{N} \rangle$.

\newpage

\begin{center}
Table 1
\vspace{1cm}
\end{center}
\begin{center}
\begin{tabular}{|c|c|c|c|} \hline\centering
& & &\\
${\em Labels}$ & \multicolumn{1}{c|}{{\em Interaction}} & \multicolumn{1}{c|}
{{\em Coupling constant}}&\multicolumn{1}{c|}{{\em Coupling constant}}\\
&&{\em in the ansatz} &{\em in the hamiltonian $H^{FP}$ }\\
& & &\\
\hline\hline
& & &\\
$I_{1}=(0,0,0,0)$& $\emptyset$ & $\emptyset$&$\emptyset$\\
& & &\\ \hline
&& &\\
$I_{2}=(1,1,0,0)$&&&\\
&&&\\
$I_{3}=(0,1,1,0)$&&&\\
&$\sigma_{j}^{z}\sigma_{j+1}^{z}$&$\alpha_{1}$&$\lambda$\\
$I_{4}=(0,0,1,1)$&&&\\
&&&\\
$I_{5}=(1,0,0,1)$&&&\\
&&&\\
\hline
&&&\\
$I_{6}=(1,0,1,0)$&&&\\
&$\sigma_{j}^{z}\sigma_{j+2}^{z}$&$\alpha_{2}$&$0$\\
$I_{7}=(0,1,0,1)$&&&\\
&&&\\
\hline
&&&\\
$I_{8}=(1,1,1,1)$& $\sigma_{1}^{z}\sigma_{2}^{z}\sigma_{3}^{z}\sigma_{4}^{z}$
 & $\alpha_{3}$&$0$\\
&&&\\
 \hline\hline
\end{tabular}
\end{center}

\end{document}